\documentclass[a4paper, 11pt]{article}
\RequirePackage[OT1]{fontenc}
\usepackage{amsfonts,amssymb,graphics,epsfig,verbatim,bm,latexsym,amsmath,url,amsbsy,multirow,rotating,enumerate,graphicx,pgf,tikz,subfig,color,xcolor}
\usepackage[textheight=722pt,textwidth=500pt,centering,headheight=50pt,headsep=12pt,footskip=12pt]{geometry}
\usepackage[colorlinks,citecolor=blue,linkcolor=purple!60!black,urlcolor=purple!60!black]{hyperref}
\usepackage[framemethod=TikZ]{mdframed}
\usetikzlibrary{positioning}
\usetikzlibrary[shapes.misc]
\usetikzlibrary{arrows,automata}
\allowdisplaybreaks
\usepackage{listings}
\begin{document}
\begin{center}
\textbf{{\LARGE{Regression and Causality}}}
\end{center}\vspace*{0.5cm}

{\Large
\begin{center}
Michael Schomaker\footnote{University of Cape Town,
Centre for Infectious Disease Epidemiology and Research; Cape Town, South Africa, \href{mailto:michael.schomaker@uct.ac.za}{michael.schomaker@uct.ac.za}\\
\indent UMIT - University for Health Sciences, Medical Informatics and Technology,
Institute of Public Health, Medical Decision Making and Health
Technology Assessment; Hall in Tirol, Austria, \href{mailto:michael.schomaker@umit.at}{michael.schomaker@umit.at}
}
\end{center}
}

\begin{abstract}
The causal effect of an intervention (treatment/exposure) on an outcome can be estimated by: i) specifying knowledge about the data-generating process; ii) assessing under what assumptions a target quantity, such as for example a causal odds ratio, can be identified given the specified knowledge (and given the measured data); and then, iii) using appropriate statistical estimation techniques to estimate the desired parameter of interest. As regression is the cornerstone of statistical analysis, it seems obvious to ask: is it appropriate to use estimated regression parameters for causal effect estimation? It turns out that using regression for effect estimation is possible, but typically requires more assumptions than competing methods. This manuscript provides a comprehensive summary of the assumptions needed to identify and estimate a causal parameter using regression and, equally important, discusses the resulting implications for statistical practice.
\end{abstract}

\section{Introduction}\label{sec:introduction}
Regression is one of the most popular statistical methods in science. The search term ``regression'' gives 1,164,884 hits on \textit{Scopus} and 793,309 hits on \textit{Web of Science}\footnote{as of 22 August 2018}. Regression models serve many purposes: they can be used for prediction, describe relationships in the data, be auxiliary tools to facilitate other estimation techniques, and they can -- in principle -- also be used to assess the effect of an intervention on an outcome.

The goals of many scientific questions are causal in nature: does a new social grant policy decrease unemployment? Will a new drug lead to lower 5-year mortality compared to the currently used drug? Is the implementation of a new advertisement strategy going to increase future sales? Does the use of a certain pesticide increase the risk of developing a specific cancer? The list is endless. No matter what study design: to answer causal questions, researchers often fit a regression model containing the outcome, the intervention and covariates, and the association between intervention and outcome is reported as the effect measure of interest. However, as Hernan (2018) highlights: ``An association measure from an observational analysis may be a biased estimate of a causal effect, but being explicit about the goal of the analysis is a prerequisite for good science. Do we want to estimate the association measure or the causal effect measure?'' \cite{Hernan:2018}. If the goal of a study is causal, and there is knowledge about the data and the process generating it, one can ask whether regression models can be used to estimate a particular causal effect or not.

In theory, regression models can identify causal effects, both for randomized studies and observational data, though the assumptions are particularly strict for the latter. This has been acknowledged in the literature, but educational articles typically focus on one aspect that may bias effect estimates, but not on a full set of assumptions. For example, Cole et al. demonstrate that the inclusion of variables which are caused by both the outcome and intervention (colliders) can cause bias \cite{Cole:2010}; Petersen et al. evaluate the impact of positivity violations, i.e. that the probability of treatment assignment is (close to) zero in certain covariate strata \cite{Petersen:2012}; and Hernan et al. point out the possible bias of per-protocol and intention-to-treat analyses in randomized trials \cite{Hernan:2012}.

While the limitations of regression modeling for causal effect estimation are well-known, a comprehensive summary of relevant assumptions is missing in the literature so far. This manuscript aims at filling this gap and provides comprehensive discussions on various aspects of regression analysis, including (causal) interpretation of regression coefficients, considerations for model selection and appropriate setups for Monte-Carlo simulations. The manuscript is structured in two parts: the first part (Sections 2 and 3) reviews the assumptions needed to identify and estimate causal parameters with regression models. This review is based on a large set of assumptions, all of which are discussed in the literature; but this set is later distilled into a smaller set given the overlap of many of those assumptions. The second part (Section 4-7) reflects on the practical implications for regression analyses: given the set of assumptions, what are the implications for different study designs, how does it affect interpretations, what are the consequences when designing simulation setups, and how should variables be selected?

\section{Framework}\label{sec:framework}

\subsection{Notation}
Let $Y$ denote the vector of outcomes, $A$ an intervention of interest, $\mathbf{L}=\{L^1,\ldots,L^p\}$ a set of measured covariates, and $\mathbf{U}=\{U^1,\ldots,U^q\}$ a set of unmeasured covariates. Using a counterfactual framework, one can think of potential outcomes $Y^{A=a}_i = Y^a_i$ as the outcome that would have been observed for individual $i$ under the intervention $A=a$, possibly contrary to the fact. If $A$ is binary, each unit has two potential outcome, $Y^1_i$ and $Y^0_i$, and only one of them can be observed. Similarly, $L^{j,a}_i$ would denote the counterfactual covariate $L^j$ for observation $i$ under $A=a$. For simplicity, longitudinal data is not considered in this paper. Notation regarding censoring, missing data and sample selection is introduced further below.

\subsection{Target Parameter}\label{sec:framework_target_parameter}
If $A$ is binary, a common estimand of interest would be the average treatment effect (ATE):
\begin{eqnarray}\label{eqn:ATE}
\text{ATE} &=& E(Y^1) - E(Y^0) \,.
\end{eqnarray}
The ATE compares the expected value of $Y$ if \textit{every} unit had received $A=1$ compared to if \textit{every} unit had received $A=0$. For a binary outcome one can, for example, look at the marginal (causal) odds ratio:
\begin{eqnarray}\label{eqn:MOR}
\text{MOR} &=& { \frac{P(Y^1=1)}{1-P(Y^1=1)} }/{\frac{P(Y^0=1)}{1-P(Y^0=1)}} \,.
\end{eqnarray}
More generally, for example when $A$ is continuous, one could specify a marginal structural working model (MSM) which describes the relationship between the intervention and the \textit{counterfactual} outcome. For example,
\begin{eqnarray}
E(Y^a) &=& \beta_0 + \beta_1 A
\end{eqnarray}
is a working MSM. Now, let $F_{\mathcal{P}}$ denote the distribution of $\mathcal{P}=\{\mathbf{L},(Y^a: a \in \mathcal{A})\}$ where $\mathcal{A}$ denotes the possible values $A$ can take; then a more general
MSM, possibly conditional on one (or more) covariate(s) $\mathbf{L}^{\ast}$, can be written as
\begin{eqnarray}\label{eqn:MSM}
E_{F_{\mathcal{P}}}(Y^a|\mathbf{L}^{\ast}) &=& m(a,\mathbf{L}^{\ast}|\beta)\,.
\end{eqnarray}
The functional form of the MSM will typically not be known. One may assume that the functional form in (\ref{eqn:MSM}) is correct, and if this is true then $\beta$ describes the true causal relationship between $A$ and $Y$; however, it is more likely that such a ``working'' model is incorrect in which case $\beta$ is defined as the projection of the true causal estimand $E_{F_{\mathcal{P}}}(Y^a|\mathbf{L}^{\ast})$ onto the specified model. This subtlety is however not the focus of this manuscript. The point is: no matter how a causal target quantity is defined, one can assess whether regression can, in principle, be used to estimate this quantity (if it is identified). In the below examples we focus on the ATE and MOR, but the arguments made hold for a wider class of target parameters.

Causal target quantities can be marginal or conditional with respect to $\mathbf{L}$. Equations (\ref{eqn:ATE}) and (\ref{eqn:MOR}) describe marginal estimands; equation (\ref{eqn:MSM}) a conditional estimand. Another simple conditional estimand is $E(Y^1|L^{\ast})$ where $L^{\ast}$ describes a particular covariate. The most meaningful causal estimands are either marginal, or conditional on very few strata; for example conditional on one or two covariates. For instance, if one is interested in the effect of a new drug (compared to an old drug) on 5-year mortality, then a marginal target quantity such as the ATE is most meaningful. The drug may however work differently is certain groups, such as pregnant women or people with concomitant medication. In such cases, estimates conditional on certain covariate strata can be meaningful. However, it would not make sense to then stratify such a causal estimand further; for example by conditioning on age, region of birth, marital status and so on. \textit{Ultimately, a causal estimand will in most cases be marginal with respect to at least a subset of $\mathbf{L}$. This is a very important point which becomes much clearer below.}

\subsection{Identification using Regression Models}
Target quantities, as described in the section above, are of interest whenever one is interested in the causal effect of $A$ on $Y$. Note that these quantities are typically marginal with respect to $\mathbf{L}$ and $\mathbf{U}$, or at least a subset thereof. In contrast, regression models provide estimates for \textit{conditional} expectations. For instance, a linear regression model may estimate $E(Y|A,\mathbf{L})$. This is an expression that refers to a conditional expectation of the \textit{observed} outcome, rather than the counterfactual one. The question posed in this paper is, in principal: under which assumptions is

\begin{eqnarray}\label{eqn:identification}
E(Y^a) &\stackrel{?}{=}& E(Y|A=a,\mathbf{L})
\end{eqnarray}

In terms of the ATE we could ask: when is

\begin{eqnarray}\label{eqn:identification2}
E(Y^1) - E(Y^0) &\stackrel{?}{=}& E(Y|A=1,\mathbf{L}) - E(Y|A=0,\mathbf{L})
\end{eqnarray}

which relates to estimating $\beta_1$ in a regression model of the form $E(Y|A,L) = \beta_0 + \beta_1 A + \beta_2 L$. Similarly, we could ask: when is the marginal causal odds ratio equivalent to the conditional odds ratio provided by logistic regression? Of course, such identification problems relate to any regression model, no matter whether more general model classes are considered or not; any generalized linear model, or any model that models relationships non-linearly, say with splines, is still a regression model that estimates a conditional expectation for an observed outcome.  Section \ref{sec:assumptions} delves deeper into the question of identification.

\subsection{Structural/Causal Model}
\begin{figure}[ht!]
\begin{center}
\subfloat[]{\label{figure:DAGa}
\begin{tikzpicture}
\node[text centered] (t) {$L$};
\node[right = 1.5 of t, text centered] (m) {$A$};
\node[right=1.5 of m, text centered] (y) {$Y$};
\draw[->, line width= 1] (t) -- node[above,font=\footnotesize]{}  (m);
\draw [->, line width= 1] (m) -- node[above,font=\footnotesize]{}  (y);
\draw[->, line width=1] (t) to  [out=45,in=135, looseness=0.5] node[below,font=\footnotesize]{} (y);
\end{tikzpicture}
}\hfill
\subfloat[]{\label{figure:DAGb}
\begin{tikzpicture}
\node[text centered] (t) {$A$};
\node[right = 1.5 of t, text centered] (m) {$Y$};
\node[right=1.5 of m, text centered] (y) {$L$};
\draw [->, line width= 1] (m) -- node[above,font=\footnotesize]{}  (y);
\draw[->, line width=1] (t) to  [out=45,in=135, looseness=0.5] node[below,font=\footnotesize]{} (y);
\end{tikzpicture}
}\hfill
\subfloat[]{\label{figure:DAGc}
\begin{tikzpicture}
\node[text centered] (z1) {$L_1$};
\node[right = 1.5 of z1, text centered] (z3) {$L_2$};
\node[right=1.5 of z3, text centered] (z2) {$U$};
\node[below=1 of z1, text centered] (a) {$A$};
\node[below=1 of z2, text centered] (y) {$Y$};
\draw[->, line width= 1] (z1) -- node[above,font=\footnotesize]{}  (a);
\draw[->, line width= 1] (z1) -- node[above,font=\footnotesize]{}  (z3);
\draw[->, line width= 1] (z2) -- node[above,font=\footnotesize]{}  (z3);
\draw[->, line width= 1] (z2) -- node[above,font=\footnotesize]{}  (y);
\draw[->, line width= 1] (z3) -- node[above,font=\footnotesize]{}  (a);
\draw[->, line width= 1] (z3) -- node[above,font=\footnotesize]{}  (y);
\draw[->, line width= 1] (a) -- node[above,font=\footnotesize]{}  (y);
\end{tikzpicture}
}
\caption{Examples of different directed acyclic graphs (DAGs)} \label{figure:DAGs}
\end{center}
\end{figure}
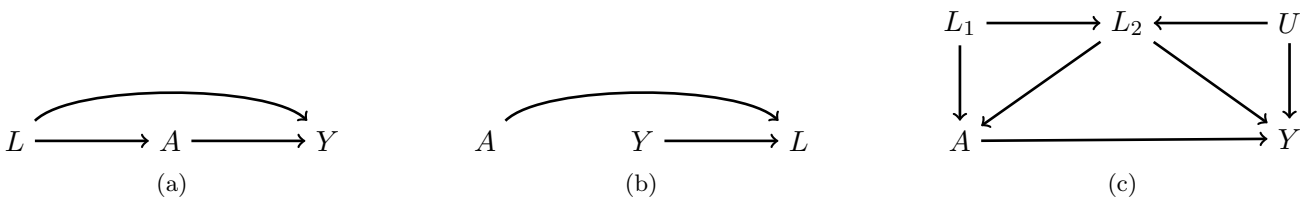

Whether identification, as outlined in (\ref{eqn:identification}), is possible depends not only on the chosen statistical estimation technique, but also on the data generating process and the measured data. Causal effect estimation can not be done without a structural model that summarizes subject matter knowledge. One way to express this knowledge is by using directed acyclic graphs (DAGs). In a DAG, each circle represents a variable -- and an arrow from A to B ($A \rightarrow B$) means that we assume that A causes B. More importantly, the absence of an arrow means we assume no causal relationship between the two respective variables. DAGs may contain both measured and unmeasured variables, see Figure \ref{figure:DAGs} for some simple examples.

There are other ways to express knowledge, such as using structural equations, but it is not necessarily needed to embark upon these approaches in more detail to explain the assumptions below. Given the structural model, the chosen target quantity and the observed data, it can be assessed whether it is possible to translate the causal question into a statistical estimand \cite{Petersen:2014}. If the answer is yes, an appropriate estimator can be chosen. We now assume that the estimation technique of interest is a regression model and ask under which assumptions this model can link the observed data to the causal target quantity.

\section{List of Assumptions}\label{sec:assumptions}
The below sections \ref{sec:assumptions1} to \ref{sec:assumptions12} summarize assumptions which are often cited to be needed to identify (and later estimate) a causal target parameter using regression coefficients. Not all of these assumptions are \textit{necessary} assumptions \cite{Greenland:2017} in the sense that in some settings they may be relaxed, or that they may even be seen as a theorem rather than an assumption \cite{Pearl:2010b}. Moreover, many of the listed assumptions overlap, and relate to similar issues; but they are nevertheless useful when practically evaluating the appropriateness of regression analysis to make causal claims. Section \ref{sec:assumptions_all} distills the list of assumptions into a smaller set and provides a formal evaluation of this reduced set. In summary, the provided list is meant to be both a comprehensive and useful guide for a given research question and analysis.

\subsection{Assumption 1: No unmeasured confounders}\label{sec:assumptions1}
The assumption of no unmeasured confounding is related to concepts of conditional exchangeability, unconfoundedness, $d$-seperation of $Y$ and $A$, closing back-door paths, randomisation assumptions, no omitted variables, ignorability of treatment assignment, selection on observables, among others. As a starting point, consider conditional exchangeability, which requires the counterfactual outcome for those treated to be the same as for those untreated if untreated subjects had received, possibly contrary to the fact, the treatment of the treated and therefore, the two groups are exchangeable (conditional on the covariate strata). Or, in other words, no matter whether the subset of those actually treated or the group of those untreated are selected, counterfactual outcomes would be the same for a given intervention $A=a$. More formally, conditional exchangeability can be expressed as
\begin{eqnarray}
Y^{{a}} \coprod {A|\mathbf{L}} \quad \text{for} \quad \forall A=a, \mathbf{L}=\mathbf{l}\,.
\end{eqnarray}
For a binary outcome and binary intervention this equates to $P(Y^a=1|A=1, \mathbf{L}=\mathbf{l})$ $=$ $P(Y^a=1|A=0, \mathbf{L}=\mathbf{l})$. To understand when this assumption is met, and how it relates to unmeasured variables, the best is to evaluate the graphical counterpart of the conditional exchangeability definition.

In a DAG, a back-door path is defined as path from $A$ to $Y$ that starts with an arrow into A. Consider Figure \ref{figure:DAGa}: here $A \leftarrow L \rightarrow Y$ is a back-door path. A back-door path generates confounding  for the effect of $A$ on $Y$, and $L$ is the confounder. A back-door path can be blocked, and confounding removed, by conditioning on the confounder\footnote{and by not conditioning on any collider, see Section \ref{sec:assumptions3}}, for example by using regression \cite{Pearl:2010}. Conditional exchangeability is the same as being able to block all back-door paths \cite{Hernan:2020}. Thus, conditional exchangeability is met if all back-door paths can be blocked by adding the respective confounders as covariates into the regression model.

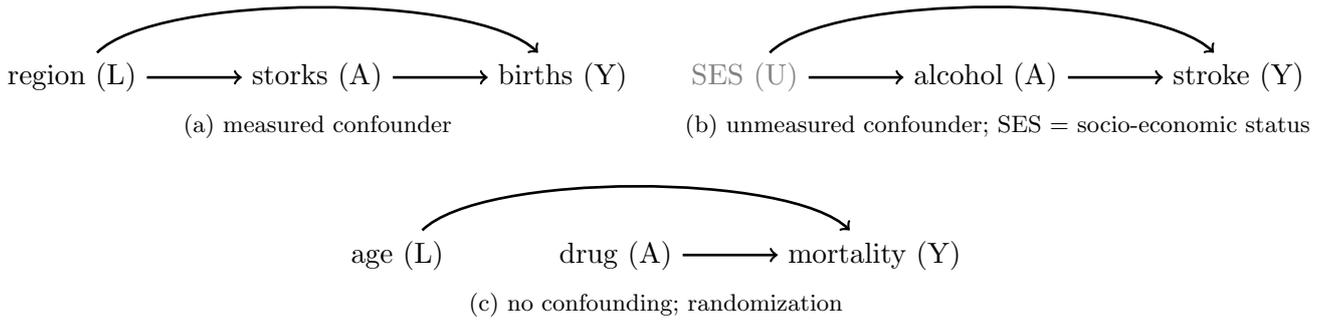
\begin{figure}[ht!]
\begin{center}
\subfloat[measured confounder]{\label{figure:DAG2a}
\begin{tikzpicture}
\node[text centered] (t) {region (L)};
\node[right = 1.25 of t, text centered] (m) {storks (A)};
\node[right=1.25 of m, text centered] (y) {births (Y)};
\draw[->, line width= 1] (t) -- node[above,font=\footnotesize]{}  (m);
\draw [->, line width= 1] (m) -- node[above,font=\footnotesize]{}  (y);
\draw[->, line width=1] (t) to  [out=45,in=135, looseness=0.5] node[below,font=\footnotesize]{} (y);
\end{tikzpicture}
}\hfill
\subfloat[unmeasured confounder;  SES = socio-economic status]{\label{figure:DAG2b}
\begin{tikzpicture}
\node[text centered] (t) {\textcolor{gray}{SES (U)}};
\node[right = 1.25 of t, text centered] (m) {alcohol (A)};
\node[right=1.25 of m, text centered] (y) {stroke (Y)};
\draw[->, line width= 1] (t) -- node[above,font=\footnotesize]{}  (m);
\draw [->, line width= 1] (m) -- node[above,font=\footnotesize]{}  (y);
\draw[->, line width=1] (t) to  [out=45,in=135, looseness=0.5] node[below,font=\footnotesize]{} (y);
\end{tikzpicture}
}\hfill
\subfloat[no confounding; randomization]{\label{figure:DAG2c}
\begin{tikzpicture}
\node[text centered] (t) {age (L)};
\node[right = 1.25 of t, text centered] (m) {drug (A)};
\node[right=1.25 of m, text centered] (y) {mortality (Y)};
\draw [->, line width= 1] (m) -- node[above,font=\footnotesize]{}  (y);
\draw[->, line width=1] (t) to  [out=45,in=135, looseness=0.5] node[below,font=\footnotesize]{} (y);
\end{tikzpicture}
}
\caption{Examples for confounding} \label{figure:DAGs-seq-exch}
\end{center}
\end{figure}

Now, if the confounder is measured, as in the example in Figure \ref{figure:DAG2a}, then regression can deal with it. It is obvious that an unmeasured confounder, as in Figure \ref{figure:DAG2b}, can not be taken into account by regression analyses; thus, there remain open back-door paths and conditional exchangeability is not achieved. In a randomized experiment, treatment assignment is random and does not depend on covariates, see Figure \ref{figure:DAG2c} for an example. The lack of an arrow from $L$ to $A$ indicates this knowledge and implies that $L$ is not a confounder. Thus, the assumption of no unmeasured confounders is achieved by design\footnote{see Section \ref{sec:obsdata_and_expdata} for details}. In this case, it will not necessarily hurt to add $L$ to the regression model\footnote{unless one is interested in non-collapsible effect measures, see Section \ref{sec:assumptions9}}, but omitting it is not wrong either.

In summary, the postulated structural model helps with identifying open back-door paths and relevant confounders. If these confounders are measured, regression adjustment is possible.

\subsection{Assumption 2: Correct model specification}\label{sec:assumptions2}
\begin{figure}[ht!]
\begin{center}
\begin{tikzpicture}
\node[text centered] (a) {$A$};
\node[right = 1. of a, text centered] (fp) {};
\node[right = 2 of a, text centered] (fp2) {};
\node[right=3 of a, text centered] (y) {$Y$};
\node[above=1.5 of fp, text centered] (l) {$L$};
\node[above=1.5 of fp2, text centered] (l2) {$L^2$};
\draw[->, line width= 1] (l) -- (a);
\draw[->, line width= 1] (l2) -- (a);
\draw[->, line width= 1] (l) -- (y);
\draw[->, line width= 1] (l2) -- (y);
\draw[->, line width= 1] (a) -- (y);
\draw[->, line width= 1] (l) -- (l2);
\end{tikzpicture}
\end{center}
\caption{A graphical representation of non-linear relationships} \label{figure:DAGs-correct-model-specs}
\end{figure}
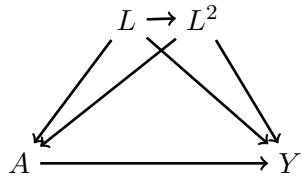

Another assumption that is required to make regression a legitimate tool for effect estimation is correct model specification, at least for observational data\footnote{in randomized trials several standard regression-based hypothesis tests (which use robust variance estimators) are guaranteed to have correct Type I error for large samples, even when the models are incorrectly specified, see Rosenblum and van der Laan for details \cite{Rosenblum:2009}}. While the causal inference literature often prominently cites this assumption, it isn't explained very often. A  simple motivation is given in Figure \ref{figure:DAGs-correct-model-specs}. Here, both $L$ and $L^2$ act as a confounder because they both affect treatment assignment and the outcome. If this is true, then there exists the back-door path $A \leftarrow L^2 \rightarrow Y$. Omitting $L^2$ from the regression equation will thus leave a back-door path open, confounding is going to persist, and the arguments made with respect to assumption 1 apply. Note that DAGs are typically completely non-parametric\footnote{A DAG may relate to a set of structural equations which which is an alternative way to encode structural assumptions and a possible way to define counterfactuals, see for example Pearl (2009)\cite{Pearl:2009}} and don't contain parametric assumptions encoded in variables such as $L^2$; Figure  \ref{figure:DAGs-correct-model-specs} is more restrictive for the sole purpose to motivate for the need to correctly specify a regression model in causal analyses.

It is important to note that the above argument concerning model mis-specification only applies to non-saturated models. A saturated model is one in which the number of parameters in the model is equal to the number
of population quantities that can be estimated by using the model. For example, consider the situation in which there is only one binary confounder $L$, as well as a binary treatment and binary outcome. A regression model $P(Y=1|A=a,L=l) = \beta_0 + \beta_1A + \beta_2L + \beta_3AL$ contains 4 parameters, and it estimates 4 quantities: $P(Y=1|A=0,L=0)$, $P(Y=1|A=1,L=0)$, $P(Y=1|A=0,L=1)$, $P(Y=1|A=1,L=1)$; thus, the model is saturated and imposes no modeling restrictions. There is no need to worry about model mis-specification here. However, as soon as $L$ is continuous and takes on many possible values, the model becomes non-saturated. In this case, correct model specification is crucial for effect estimation.

\subsection{Assumption 3: No conditioning on a collider}\label{sec:assumptions3}
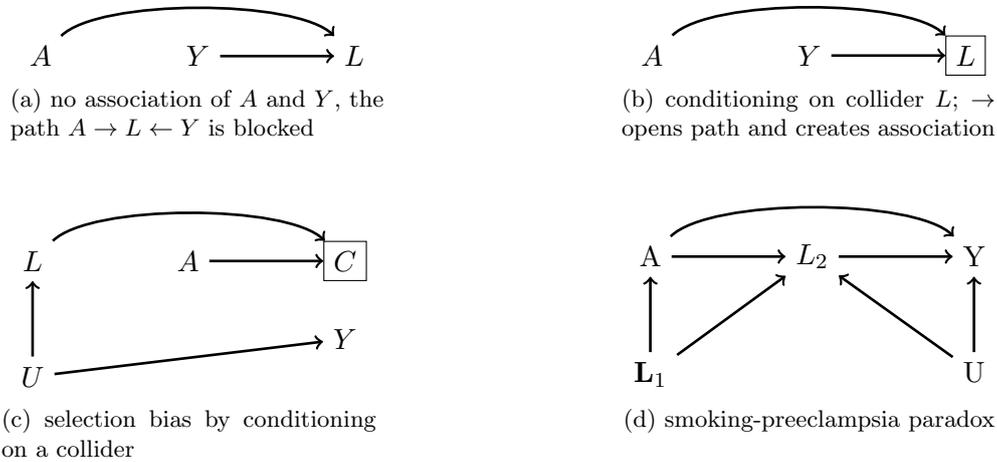
\begin{figure}[ht!]
\begin{center}
\subfloat[no association of $A$ and $Y$, the  path $A \rightarrow L \leftarrow Y$ is blocked]{\label{figure:DAG3a}
\begin{tikzpicture}
\node[text centered] (t) {$A$};
\node[right = 1.5 of t, text centered] (m) {$Y$};
\node[right=1.5 of m, text centered] (y) {$L$};
\draw [->, line width= 1] (m) -- node[above,font=\footnotesize]{}  (y);
\draw[->, line width=1] (t) to  [out=45,in=135, looseness=0.5] node[below,font=\footnotesize]{} (y);
\end{tikzpicture}
}\hspace*{3cm}
\subfloat[conditioning on collider $L$; $\rightarrow$ opens path and creates association]{\label{figure:DAG3b}
\begin{tikzpicture}
\node[text centered] (t) {$A$};
\node[right = 1.5 of t, text centered] (m) {$Y$};
\node[draw, right=1.5 of m, text centered] (y) {$L$};
\draw [->, line width= 1] (m) -- node[above,font=\footnotesize]{}  (y);
\draw[->, line width=1] (t) to  [out=45,in=135, looseness=0.5] node[below,font=\footnotesize]{} (y);
\end{tikzpicture}
}\\[0.3cm]
\subfloat[selection bias by conditioning on a collider]{\label{figure:DAG3c}
\begin{tikzpicture}
\node[text centered] (l) {$L$};
\node[right = 1.5 of l, text centered] (a) {$A$};
\node[draw,right=1.5 of a, text centered] (c) {$C$};
\node[below=0.5 of c, text centered] (y) {$Y$};
\node[below=1 of t, text centered] (u) {$U$};
\draw[->, line width= 1] (a) -- node[above,font=\footnotesize]{}  (c);
\draw[->, line width=1] (l) to  [out=45,in=135, looseness=0.5] node[below,font=\footnotesize]{} (c);
\draw[->, line width= 1] (u) -- node[above,font=\footnotesize]{}  (l);
\draw[->, line width= 1] (u) -- node[above,font=\footnotesize]{}  (y);
\end{tikzpicture}
}\hspace*{3cm}
\subfloat[smoking-preeclampsia paradox]{\label{figure:DAG3d}
\begin{tikzpicture}
\node[text centered] (R) {A};
\node[below=1 of R, text centered,align=center] (bmi) {$\mathbf{L}_1$};
\node[right=1.5 of R, text centered,align=center] (sw) {$L_2$};
\node[right=1.5 of sw, text centered,align=center] (y) {Y};
\node[below=1 of y, text centered] (pp) {U};
\draw[->, line width= 1] (R) -- (sw);
\draw[->, line width= 1] (bmi) -- (R);
\draw[->, line width= 1] (pp) -- (sw);
\draw[->, line width= 1] (pp) -- (y);
\draw[->, line width= 1] (bmi) -- (sw);
\draw[->, line width= 1] (sw) -- (y);
\draw[->, line width=1] (R) to  [out=45,in=135, looseness=0.5](y);
\end{tikzpicture}
}
\caption{Examples for colliders} \label{figure:DAGs-collider}
\end{center}
\end{figure}

As highlighted under assumption 1, any back-door path between $A$ and $Y$ needs to be blocked to identify the desired target quantity. Conditioning on a confounder blocks a path, but conditioning on a so-called \textit{collider} opens a path. A variable is a collider on a given path if the path enters and leaves the respective variable via arrowheads. In Figure \ref{figure:DAG3a}, $L$ is a collider. No adjustment is needed, the  path $A \rightarrow L \leftarrow Y$ is blocked. Conditioning on $L$, for example via regression, opens the path and introduces association where there was none before. This situation is illustrated in Figure \ref{figure:DAG3b} where the rectangle means `conditioning on'.

The bias introduced by conditioning on a common effect of treatment and outcome is not only known as collider bias, but sometimes also as selection bias (see also Section \ref{sec:assumptions_sample}). Consider Figure \ref{figure:DAG3c} in which $A$ refers to a meal containing wasabi, $C$ an indicator whether a respondent agreed to take part in the survey, $L$ the stress level, $U$ an unmeasured heart disease and $Y$ mortality 1 year after the survey. Conditioning on $C$ is another expression for saying that the data may be analyzed on the complete cases only, i.e. restricting the analysis to those respondents who agreed to take part in the survey. It may not be surprising to find out that those who disagreed may not be representative of all people approached, and therefore there is selection bias. In this (hypothetical) example conditioning on $C$ opens the path $A \rightarrow C \leftarrow L \leftarrow U \rightarrow Y$ and introduces selection bias in the sense that those included in the study have a lower stress level and may therefore be less likely to die during the following year. If the unmeasured heart disease was measured, the path could be blocked again and collider bias would not occur.

Of course, Figure \ref{figure:DAG3c} is an over-simplified example omitting many important variables. A somewhat more realistic example is represented in Figure \ref{figure:DAG3d}. Here, $A$ refers to the amount of smoking, $Y$ refers to pre-eclampsia diagnosed at birth (a common birth complication), $\mathbf{L}_1$ represents the BMI and age of the mother as well as the number of prior births, $L_2$ is the gestation week at delivery and $U$ is the placental pathology. Multiple epidemiological studies using a regression model containing $\mathbf{L}_1$, $L_2$, and $L_3$ as covariates have concluded that higher smoking is associated with a lower risk of pre-eclampsia. This nonintuitive finding can be explained by collider bias\footnote{it is also partly due to bias because of conditioning on a mediator, see Section \ref{sec:assumptions4}} as conditioning on $L_2$ opens the path $A \leftarrow \mathbf{L}_1 \rightarrow L_2  \leftarrow U \rightarrow Y$, see Luque-Fernandez et al. (2016) for details \cite{Luque:2016}.

Other popular examples of collider biases are the obesity paradox \cite{Banack:2014}, the birthweight paradox \cite{Whitcomb:2009}, the sodium intake paradox \cite{Luque:2019}, and survival bias \cite{Schomaker:2019}, among others.

In summary, regression models should not contain colliders which open a back-door path from $A$ to $Y$.

\subsection{Assumption 4: No conditioning on a mediator (and its descendants)}\label{sec:assumptions4}
\begin{figure}[ht!]
\begin{center}
\subfloat[Conditioning on the mediator blocks the path $A \rightarrow M \rightarrow Y$ and introduces bias with respect to the total effect of $A$ on $Y$]{\label{figure:DAG_med_a}
\begin{tikzpicture}
\node[text centered] (a) {$A$};
\node[draw,right = 1.5 of t, text centered] (m) {$M$};
\node[right=1.5 of m, text centered] (y) {$Y$};
\draw [->, line width= 1] (a) -- node[above,font=\footnotesize]{}  (m);
\draw [->, line width= 1] (m) -- node[above,font=\footnotesize]{}  (y);
\draw[->, line width=1] (a) to  [out=45,in=135, looseness=0.5] node[below,font=\footnotesize]{} (y);
\end{tikzpicture}
}\hspace*{3cm}
\subfloat[Conditioning on $L$, a post-treatment variable, is fine here]{\label{figure:DAG_med_b}
\begin{tikzpicture}
\node[text centered] (a) {$A$};
\node[draw,right = 1.5 of a, text centered] (l) {$L$};
\node[right=1.5 of l, text centered] (y) {$Y$};
\node[above = 0.5 of a, text centered] (u) {$U$};
\draw [->, line width= 1] (u) -- node[above,font=\footnotesize]{}  (a);
\draw [->, line width= 1] (u) -- node[above,font=\footnotesize]{}  (m);
\draw [->, line width= 1] (m) -- node[above,font=\footnotesize]{}  (y);
\draw[->, line width=1] (a) to  [out=45,in=135, looseness=0.5] node[below,font=\footnotesize]{} (y);
\end{tikzpicture}
}\\[0.3cm]
\subfloat[Conditioning on a descendent of a mediator is problematic]{\label{figure:DAG_med_c}
\begin{tikzpicture}
\node[text centered] (a) {$A$};
\node[right = 1.5 of t, text centered] (m) {$M$};
\node[right=1.5 of m, text centered] (y) {$Y$};
\node[draw,below = 0.5 of m, text centered] (l) {$L$};
\draw [->, line width= 1] (a) -- node[above,font=\footnotesize]{}  (m);
\draw [->, line width= 1] (m) -- node[above,font=\footnotesize]{}  (l);
\draw [->, line width= 1] (m) -- node[above,font=\footnotesize]{}  (y);
\draw[->, line width=1] (a) to  [out=45,in=135, looseness=0.5] node[below,font=\footnotesize]{} (y);
\end{tikzpicture}
}\hspace*{3cm}
\subfloat[Conditioning on $L$ is possible]{\label{figure:DAG_med_d}
\begin{tikzpicture}
\node[text centered] (a) {$A$};
\node[right = 1.5 of t, text centered] (m) {$M$};
\node[right=1.5 of m, text centered] (y) {$Y$};
\node[draw,below = 0.5 of m, text centered] (l) {$L$};
\draw [->, line width= 1] (a) -- node[above,font=\footnotesize]{}  (m);
\draw [->, line width= 1] (l) -- node[above,font=\footnotesize]{}  (m);
\draw [->, line width= 1] (m) -- node[above,font=\footnotesize]{}  (y);
\draw[->, line width=1] (a) to  [out=45,in=135, looseness=0.5] node[below,font=\footnotesize]{} (y);
\end{tikzpicture}
}
\caption{Examples for the role of mediators} \label{figure:DAG_med}
\end{center}
\end{figure}
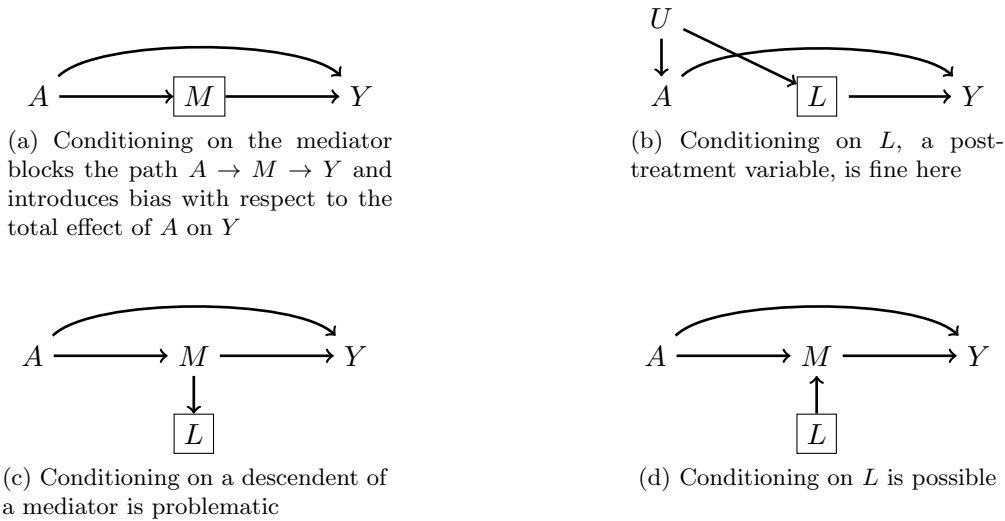

A mediator is a variable $M$ which lies on the path between $A$ and $L$, i.e. $A \rightarrow M \rightarrow Y$ as in Figure \ref{figure:DAG_med_a}. Conditioning on a mediator blocks the path which relates to the indirect effect of $A$ on $Y$ through $M$, and thus introduces bias with respect to the total\footnote{sometimes only the direct effect $A \rightarrow Y$ or only the indirect effect $A \rightarrow M \rightarrow Y$ are of interest to a researcher; however, in the context of this paper, only total effects --estimated via regression models-- are considered.} effect of $A$ on $Y$. This follows directly from both Pearl's back-door criterion \cite{Pearl:2009} and his $d$-separation concept\footnote{the latter is explained in Section \ref{sec:assumptions_all}}. The \textit{back-door criterion} states that a set $\mathcal{S}$ is sufficient for adjustment if i) the elements of $\mathcal{S}$ block all back-door paths, as explained in Section \ref{sec:assumptions1}, and ii) $\mathcal{S}$ does not contain a descendant of $A$.

In Figures \ref{figure:DAG_med_a} and \ref{figure:DAG_med_c}, conditioning on $M$ and $L$ respectively would thus introduce bias, whereas conditioning on $L$ in Figure \ref{figure:DAG_med_d} does not cause any problems \cite{Schisterman:2009}. It has been argued that, as a practical rule of thumb, one should simply not condition on post-treatment variables to avoid the problem of conditioning on mediators \cite{Imbens:2019}. However, as pointed out by others, this recommendation is not generally valid \cite{Hernan:2020}. A simple example is given in Figure \ref{figure:DAG_med_b} where conditioning on the post-treatment variable $L$ does not close any mediating path, but helps to block the back-door path $A \leftarrow U \rightarrow L \rightarrow Y$.

In summary, regression models should not contain mediators, or any of the mediator's descendants (for analyses where the total effect of $A$ on $Y$ is of interest).

\subsection{Assumption 5: Positivity}\label{sec:assumptions5}
The assumption of positivity says that the probability of treatment assignment is greater than zero in all confounder strata, i.e.
\begin{eqnarray*}
  P(A=a|\mathbf{L}=\mathbf{l})>0 \quad \text{for} \quad \forall \mathbf{l} \quad \text{with} \quad P(\mathbf{L}=\mathbf{l})\neq 0.
\end{eqnarray*}
In a randomized study this assumption is typically fulfilled because treatment assignment is probabilistic, $0 < P(A_i=1) < 1$, and confounders are not present by design (assumption 1); see Section \ref{sec:obsdata_and_expdata} for more details. In other studies the assumption of positivity could be violated for two reasons:
\begin{enumerate}[i)]
\item it may not be possible to receive $A=a$ in $\mathbf{L}=\mathbf{l}$ by definition, or
\item $P(A=a|\mathbf{L}=\mathbf{l})$ may be zero, or close to zero, in a finite sample.
\end{enumerate}
For example, in Figure \ref{figure:DAG2a} the probability to find $>1$ storks in a particular region $L=l$ may be zero, simply because there live no storks in this region. It would however be more commonly the case that the positivity assumption is practically violated. In settings where there are multiple continuous confounders it is likely that the assumption is violated, at least to some degree.

Diagnosing and responding to practical positivity violations is a topic that is heavily under-researched. In simple settings, where there is a limited set of discrete confounders and a binary treatment, one could simply produce several tables that tabulate the treatment against the different confounder strata, see Westreich and Cole (2010) \cite{Westreich:2010} and the references therein for some examples. However, in realistic settings with many confounders, possibly some of them being continuous, this approach becomes unfeasible. Alternatively, if the treatment is binary, one could estimate the propensity scores $P(A=1|\mathbf{L=l})$ and $P(A=0|\mathbf{L=l})$  in the data and evaluate the estimated probabilities \cite{Luque:2018}; if they are small, say $<0.01$, then this is likely indicating positivity violations. Unfortunately this approach won't work for a continuous $A$, and if certain covariate combinations are not present in the data practical positivity violations may remain uncovered. There are also other options, like using the parametric bootstrap \cite{Petersen:2012}, but practical positivity violations are likely going to occur in many data sets of moderate to small size with multiple confounders.

Addressing positivity violations can be challenging, but it is good to know that the positivity assumption can often be relaxed if the modeling approach is flexible; that is, if there are sparse regions of the data, i.e. covariate regions for which $P(A=a|\mathbf{L}=\mathbf{l}) \approx 0 $, then the model may be able to extrapolate to these regions if its complexity is sufficient. It is difficult to say how complex a regression model should be, but semi-parametric approaches using (penalized) splines may be a good start \cite{Wood:2006}. Nevertheless, moderate positivity violations can be best addressed by using approaches other than regression \cite{Petersen:2012}, such as using the g-formula \cite{Daniel:2011} or double robust approaches \cite{Schomaker:2019, vanderLaan:2011}. If positivity violations are severe not much can be done and changing the question of interest, or the target population, are alternatives \cite{Petersen:2012}.

In summary, positivity violations often occur  in practice. If the data do not come from randomized studies, flexible modeling strategies, and estimation approaches other than regression, are the best ways to respond to moderate positivity violations, though further research on this topic is needed.

\subsection{Assumption 6: Consistency}\label{sec:assumptions6}
The assumption or theorem of consistency is formally defined as
\begin{eqnarray}
\text{If} \, A=a, \,\text{then}\, Y^a=Y \quad \text{for}\,\forall a \,.
\end{eqnarray}
If $A$ is binary, then this is equivalent to $Y_i = A_i Y_i^1 + (1-A_i) Y_i^0$ for each observation $i$, $i=1,\ldots,n$. It seems trivial to require $Y$ to be $Y^a$ if $A=a$, but if the intervention $A=a$ is not well-defined, then the link between actual treatment and counterfactual outcome may not be clear. For example, suppose the intervention of interest is ``single parent''; then, it is not clear what single parent means. It could be someone taking care of a child on his/her own for 1 month, or for a period of 18 years, by turns with a former partner, or with support from a grandmother. In each case the potential outcome $Y^a$, let's say final graduation mark at age 18, may be different; and therefore the effect of $A=a$ on the outcome may be different for different versions of the treatment. Any intervention that can be interpreted and implemented in different ways is potentially prone to bias arising from violation of the consistency assumption.

An alternative view of this issue is that the knowledge of different treatment versions should already be captured in the DAG: then both the intervention and the target quantity could be defined more precisely. In this case the consistency assumption would be simply a theorem \cite{Pearl:2010b}.

Another scenario where the consistency assumption is violated is when observational units can influence each other, i.e. when $Y^a_i$ depends on $A_j$. This is immediately clear from the consistency definition for binary treatments. An example would be the effect of vaccinations in small geographical areas, where the experience with treatment $A=a$ of person $j$ could influence the outcome $Y^a$, such as a pain score, of person $i$; or if the effect of different pesticides is compared among neighbouring fields that are not fully separated in terms of pesticide use.

Randomized trials are less prone to violations of the consistency assumption because the trial's protocol should be clear about the exact implementation of the intervention. In observational studies it is important to unambiguously define the intervention of interest. More examples around the interpretation of the consistency assumption and its practical relevance are given in the literature \cite{Cole:2009, Rehkopf:2016, Hernan:2008}.

\subsection{Assumption 7: No interference}\label{sec:assumptions7}
Often ``no interference'' or the ``stable-unit-treatment-value assumption'' (SUTVA) is stated as an additional assumption to identify a causal target quantity. Rubin (1986) \cite{Rubin:1986} states\footnote{the notation is changed to match the notation of this paper} ``SUTVA is the assumption that the value of $Y$ for unit $i$ when exposed to treatment $a$ will be the same no matter what mechanism is used to assign treatment $a$ to unit $i$ and no matter what treatments the other units receive, and this holds for $i=1,\ldots,n$ and $\forall a \in \mathcal{A}$. SUTVA is violated when, for example, there exist unrepresented versions of treatment ($Y^a_i$ depends on which version of treatment $a$ was received) or interference between units ($Y^a_i$ depends on whether unit $j$ received treatment $a$ or $a'$)''. While the definitions of consistency and SUTVA are technically not  identical, they very much imply the same things, namely problems in identification of the causal target quantity if there exist multiple versions of treatment or if there is the possibility that observational units influence each other. In that respect consistency and SUTVA can be seen as very similar assumptions, with similar implications.

\subsection{Assumption 8: No relevant effect modification}\label{sec:assumptions8}
As opposed to other estimation techniques, regression typically requires the effect to be constant across the confounder strata, i.e. regression can identify \textit{many} causal effects only under the assumption of no relevant effect modification. To understand why effect modification is such a crucial point, it is good to understand first that based on conditional exchangeability (assumption 1) and consistency and SUTVA (assumptions 5 and 6) we can write the following in the case of a single confounder $L$:
\begin{eqnarray*}
E(Y^a|L=l) &\stackrel{A1}{=}& E(Y^a|A=a,L=l)\\
&\stackrel{A5,A6}{=}& E(Y|A=a,L=l)
\end{eqnarray*}
A linear regression model estimating $E(Y|A,L)$ would be $E(Y|A,L)$ $=$ $\beta_0 + \beta_1 A + \beta_2 AL + \beta_3 L$. For $A=1$ and $A=0$ this yields
\begin{eqnarray*}
E(Y|A=1,L=l) &=& \beta_0 + \beta_1  + \beta_2 l +\beta_3 l \,,\\
E(Y|A=0,L=l) &=& \beta_0 + \beta_3 l\,.
\end{eqnarray*}
Now suppose the ATE is the target quantity of interest. Using assumptions 1, 5 and 6 as above we can write
\begin{eqnarray*}
E(Y^1 - Y^0|L=l) &\stackrel{A1, A5, A6}{=}& E(Y|A=1,L=l) - E(Y|A=0,L=l) \\
&=& \beta_0 + \beta_1  + \beta_2 l+ \beta_3 l - \beta_0 - \beta_3 l \\
&=& \beta_1  + \beta_2 l
\end{eqnarray*}
This implies that only for constant effects, i.e. when $\beta_2=0$, the ATE [that is, $E(Y^1-Y^0)$] can be estimated using $\beta_1$. If the effect is not constant, i.e. $\beta_2\neq0$, then no \textit{marginal} causal effect will be estimated but a conditional causal effect. Certainly, as indicated in equation (\ref{eqn:MSM}), conditional effects may sometimes be of interest; however, then the assumption that this conditional effect is constant across the remaining confounder levels (if $\mathbf{L}$ is matrix with $\geq 2$ columns), still needs to hold. \textit{Adding interaction terms to regression models addresses effect modification but typically changes the target quantity from marginal quantities to conditional quantities.} As highlighted in Section \ref{sec:framework_target_parameter}, most meaningful estimands are marginal with respect to at least a subset of $\mathbf{L}$. Competing methods \cite{Hernan:2020} marginalize over the confounder distribution and don't require effects to be constant.

In summary, if effect modification is present, neglecting this in the regression model will introduce bias of causal estimands that are marginal with respect to (a subset of) $\mathbf{L}$, but addressing the problem by adding interaction terms changes the target quantity from a marginal to a conditional quantity.

\subsection{Assumption 9: Collapsibility}\label{sec:assumptions9}
Another assumption which is needed when using regression to estimate target quantities which are marginal with respect to at least a subset of $\mathbf{L}$, is that the target quantity chosen is collapsible. For illustration, consider 3 discrete variables $Y, A, L$, which are summarized in a $I \times J \times K$ table. \textit{Strict} collapsibility of a measure of association of $A$ and $Y$ means that this measure is constant across the strata of $L$ and that these conditional measures are equal to the marginal measure. Table \ref{tab:collapsibility}, from Greenland et al (1999) \cite{Greenland:1999}, shows a simple example for 3 binary variables.

\begin{table}[ht!]
\caption{Examples of collapsibility and non-collapsibility from Greenland et al (1999) \cite{Greenland:1999}}
\label{tab:collapsibility}
\begin{center}
 \renewcommand{\arraystretch}{1.15}
\begin{tabular}{p{0.15\textwidth}p{0.05\textwidth}p{0.05\textwidth}p{0.05\textwidth}p{0.05\textwidth}p{0.05\textwidth}p{0.05\textwidth}}
\hline
&\multicolumn{2}{c}{L=1}&\multicolumn{2}{c}{L=0}&\multicolumn{2}{c}{Marginal}\\
\cline{2-7}
&A=1&A=0&A=1&A=0&A=1&A=0\\
Y=1&0.2&0.15&0.10&0.05&0.30&0.20\\
Y=0&0.05&0.10&0.15&0.20&0.20&0.30\\
\cline{2-7}
&&&&&&\\
Risk$^{a}$&0.80&0.60&0.40&0.20&0.60&0.40\\
Risk difference&\multicolumn{2}{c}{0.2}&\multicolumn{2}{c}{0.2}&\multicolumn{2}{c}{0.2}\\
Risk ratio&\multicolumn{2}{c}{1.33}&\multicolumn{2}{c}{2.00}&\multicolumn{2}{c}{1.50}\\
Odds ratio&\multicolumn{2}{c}{2.67}&\multicolumn{2}{c}{2.67}&\multicolumn{2}{c}{2.25}\\
\hline
\multicolumn{7}{l}{$^{a}$ defined as probability of Y=1}
\end{tabular}
\end{center}
\end{table}

If the relationship between $A$ and $Y$ are measured by means of the risk difference, i.e. the differences in the probabilities $P(Y=1|A)$, then the conditional and marginal estimates are identical. The risk difference is therefore a measure of association that is strictly collapsible. The risk ratio is however not strictly collapsible because conditional estimates are not constant across the strata of $L$. However, while the risk ratio is not strictly collapsible, it is collapsible. A measure of association (or causal effect measure) is said to be collapsible if the marginal effect measure is equal to a weighted average of the stratum-specific associations (or causal effect measures) \cite{Huitfeldt:2019}. The marginal risk ratio lies in this case between the maximum and minimum conditional risk ratio \cite{Hernan:2020}. The odds ratio is known to be neither collapsible nor strictly collapsible.

It is not immediately evident which measures are (strictly) collapsible and if yes under what assumptions, particularly in the context of regression coefficients: here, collapsibility (for a generalized linear model) means that $\beta_1 = \beta^{\ast}_1$ for $g(E(Y|A,L)) = \beta_0 + \beta_1 A$ and $g(E(Y|A,L)) = \beta_0 + \beta_1^{\ast} A + \beta_2^{\ast} L$ (and thus the model adjusted for $L$ can be used for marginal effect estimation). For example, the conditional odds ratios from logistic regression will typically not be identical to the marginal odds ratios as defined in (\ref{eqn:MOR}), and marginal effect estimation with logistic regression is therefore not recommended. In fact, it is possible that the coefficient associated with the respective conditional odds ratio has a different sign than the marginal odds ratio. This difference is also sometimes being referred to as Simpson's paradox \cite{Hernan:2011}. The adjusted risk ratio obtained from Poisson regression is however collapsible \cite{Vellaisamy:2008}. It follows that for binary outcomes, the risk ratio obtained from Poisson regression is often a better choice than the odds ratio obtained from logistic regression for most questions where the target quantity is marginal with respect to $\mathbf{L}$. In time-to-event analyses, the Cox proportional hazards model produces coefficients that are not collapsible, whereas the Aalen hazards model produces collapsible estimates \cite{Martinussen:2013, Sjolander:2016}.

The above discussions make it clear that coefficients from regression models which are adjusted for $\mathbf{L}$ and are non-collapsible often can't be used to estimate many relevant (marginal) target quantities. This statement holds even when the covariates $\mathbf{L}$ are only related to the outcome, and not the intervention, and are thus no confounders \cite{Sjolander:2016}.

The conclusion is that non-collapsibility of an effect measure does not strictly mean that causal inference is impossible at all; but, similar to assumption 7, it is evident that the use of regression models will then provide effect estimates conditional on many covariates, which is most often not useful \cite{Luque:2019b}. If the purpose of the analysis is obtaining some sort of marginal effect, then the choice of an inappropriate regression model (in terms of collapsibility) prohibits effect estimation.

\subsection{Assumption 10: Compliance}\label{sec:assumptions10}
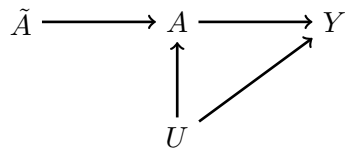
\begin{figure}[ht!]
\begin{center}
\begin{tikzpicture}
\node[text centered] (ta) {$\tilde{A}$};
\node[right = 1.5 of t, text centered] (a) {$A$};
\node[right=1.5 of a, text centered] (y) {$Y$};
\node[below=1 of a, text centered] (u) {$U$};
\draw [->, line width= 1] (ta) --  (a);
\draw [->, line width= 1] (a) --  (y);
\draw [->, line width= 1] (u) --  (y);
\draw [->, line width= 1] (u) --  (a);
\end{tikzpicture}
\caption{A simple situation of non-compliance} \label{figure:compliance}
\end{center}
\end{figure}

Non-compliance with treatment assignment is, opposed to the previous assumptions, mostly relevant in randomized studies with human subjects\footnote{it is also possible that non-compliance with treatment assignment occurs in observational studies. This is often called `measurement error', see also assumption 12.}. Here, treatment is assigned randomly to $\tilde{A}=\tilde{a}$, but the treatment actually taken by study participants ($A=a$) may differ from the assigned treatment. This situation is called non-adherence or non-compliance and is visualized in Figure \ref{figure:compliance}. It is evident that randomization is lost under non-compliance because other factors ($U$) may now affect treatment assignment, and also the outcome, and therefore confounding can occur. For example, patients may decide not to take a specific drug because they are not aware of the negative consequences involved; socio-economic factors determining this behaviour may also affect the outcome, such as a particular morbidity, and therefore adjustment for these factors would be required.

Regression models, with covariate adjustment, are often used in randomized trials to take into account sampling uncertainty and to reduce variability in the effect estimates. While often simple tests would be sufficient in an idealized trial, using regression models is not necessarily wrong\footnote{see the discussion on collapsibility though (Section \ref{sec:assumptions9}): the use of regression may change the target quantity if the effect measure is not collapsible.}. In fact, even under model mis-specification (as discussed under assumption 2) many commonly used regression models provide an asymptotically correct Type I error, at least under an i.i.d. assumption, robust variance estimation and for the most commonly employed regression models \cite{Rosenblum:2009}. However, under non-compliance these results won't hold.

If a causal quantity can be estimated under non-compliance depends on the quantity itself as well as the estimation technique.

\begin{enumerate}[i)]
\item suppose the ATE is of interest, i.e. $ E(Y^1) - E(Y^0)$. Here, the treatment's \textit{efficacy} is what matters: what is the effect of $A$ on $Y$? If $U$ in the DAG in Figure \ref{figure:compliance} is unmeasured, or if there exist other complex relationships that open a back-door path from $A$ to $Y$, then the treatment effect is not identified by fitting a regression model. Other approaches that evaluate the counterfactual intervention $\tilde{A}=A=a$\footnote{under the assumption that both $A$ and $\tilde{A}$ are measured}, like the g-formula \cite{Snowden:2011}, are needed here.
\item Given the difficulty to estimate the treatment's efficacy, it is common to evaluate the treatment's clinical \textit{effectiveness} instead. This would be the effect of $\tilde{A}=\tilde{a}$ on $Y$ and is called ``intention-to-treat'' (ITT) effect (which can be estimated from the data\footnote{an estimand similar to the ITT estimand can be estimated using instrumental variable estimation, see Hernan and Robins \cite{Hernan:2020} for details, as well as assumptions involved.}, see also Table \ref{tab:compliance}). It is obvious that $\tilde{A}=\tilde{a}$ is randomized and thus a test (or regression) can be used to estimate its (causal) effect on $Y$. It follows that for this estimand the use of regression is generally valid. The ITT approach may be useful in placebo-controlled trials, because the ITT effect can be seen as a conservative estimate of the true treatment effect (of $A$), i.e. one that is biased towards the null; but in many other settings, for example where one wants to prove a treatment's harmful effect, this is not useful because an analysis that is biased towards the null may fail to establish the harmful nature of an effect \cite{Hernan:2012}.
\item other estimands under non-adherence are the per-protocol, as treated and complier average causal effect (CACE), see Table \ref{tab:compliance}. The as-treated effect estimates the difference in expectation of $Y$ for those actually taking treatment ($A=1$) and those not ($A=0$). As indicated above, since $A$ is not randomized the as treated effect won't estimate the ATE unless other strict assumptions (i.e.  assumptions 1-8) are met. The per-protocol effect looks at the subset of those subjects that adhered to the assigned treatment. This subset may be subject to selection bias, and may not be representative of the study population, unless selection into the study ($C=1$, i.e. $A=\tilde{A}=1$) is completely ignorable, i.e. $P(C=1|Y, \mathbf{L}, U, \xi) = P(C=1| \xi)$, or ignorable ($P(C=1|Y, \mathbf{L}, U, \xi) = P(C=1|Y, \mathbf{L}, \xi)$) and being accounted for; see assumption 11 and Section \ref{sec:assumptions_sample} for more details. The CACE, which measures the treatment effect among those who comply with their assigned treatment, is always greater than the ITT and is an alternative valid causal effect measure. The CACE can identify the ATE when the ATE is the same for compliers and non-compliers, if they had in fact complied \cite{Little:2009}.
\end{enumerate}

\begin{table}[ht!]
\caption{Estimands that are commonly looked at under non-compliance for a binary intervention}
\label{tab:compliance}
\begin{center}
 \renewcommand{\arraystretch}{1.25}
\begin{tabular}{p{0.24\textwidth}|p{0.52\textwidth}}
effect & often estimated using \\
\hline
intention-to-treat (ITT) & $P(Y=1|\tilde{A}=1)- P(Y=1|\tilde{A}=0)$ \\
as treated& $P(Y=1|A=1)-P(Y=1|A=0)$\\
per protocol& $P(Y=1|A=1,\tilde{A}=1)-P(Y=1|A=0,\tilde{A}=0)$\\
CACE  & $[P(Y=1|\tilde{A}=1)-P(Y=1|\tilde{A}=0)]/{P(A=1|\tilde{A}=1)}$\\
\end{tabular}
\end{center}
\end{table}

In summary, using regression under non-adherence will typically yield biased effect estimates with respect to the ATE and other common marginal effect measures. An alternative is re-defining the estimand as the ITT or CACE, though this redefines the question of interest and may not be advisable in studies that are designed to quantify harmful effects or in non-inferiority studies \cite{Hernan:2012}.

\subsection{Assumption 11: the missing data mechanism}\label{sec:assumptions11}
If missing data causes bias with respect to the (causal) target quantity ultimately depends on the the mechanism that gives rise to the missing data and the target quantity itself.

Traditionally, definitions such as missing completely at random (MCAR), missing at random (MAR), and missing not at random (MNAR) have been used to determine if the missingness mechanism can be ignored or methodologically addressed. Let $D=\{\mathbf{L}, A, Y\}$ denote the stacked data vector of length $(p+2) \times n$ and $C$ a vector of the same size with $C_i=1$ if $D_i$ is observed and $C_i=0$ otherwise. The vector $C$ partitions $D$ into the two subsets of observed and unobserved data: $D^{\text{obs}}$ with $\text{obs}=\{i:c_i=1\}$ and $D^{\text{mis}}$ with $\text{mis}=\{i:c_i=0\}$. A common definition\footnote{the given definitions are sometimes also called ``always'' \cite{Mealli:2015} or ``everywhere'' \cite{Seaman:2013} MCAR, MAR and MNAR. In some cases, weaker conditions that refer to a specific realized sample have been proposed; see Doretti et al. \cite{Doretti:2018} for an overview. A stronger, graphical definition of MAR is explained below.} \cite{Mealli:2015,Rubin:1976} of the above mentioned missingness concepts are:

\begin{eqnarray*}
\text{MCAR:} & P(C=c|D^{\text{obs}}=d^{\text{obs}}, D^{\text{mis}}=d^{\text{mis}};\xi) =&    \hspace*{-0.25cm} P(C=c|\xi) \quad\quad\quad\quad\quad\quad\quad\quad\quad\quad\quad\quad \forall \xi,c,d^{\text{obs}},d^{\text{mis}}\,,\\
\text{MAR:}  & P(C=c|D^{\text{obs}}=d^{\text{obs}}, D^{\text{mis}}=d^{\text{mis}};\xi) =&    \hspace*{-0.25cm} P(C=c|D^{\text{obs}}=d^{\text{obs}};\xi) \quad\quad\quad\quad\quad\quad\,\,\, \forall \xi,c,d^{\text{obs}},d^{\text{mis}}\,,\\
\text{MNAR:} & P(C=c|D^{\text{obs}}=d^{\text{obs}}, D^{\text{mis}}=d^{\text{mis}};\xi) \neq& \hspace*{-0.25cm} P(C=c|D^{\text{obs}}=d^{\text{obs}}, D^{\text{mis}}=d^{\text{mis}}_{\ast};\xi)\\
 && \hspace*{2.2cm}\text{for some} \quad \xi,c,d^{\text{obs}} \quad \text{and} \quad d^{\text{mis}}\neq d^{\text{mis}}_{\ast}\,.\\
\end{eqnarray*}

MCAR means that the probability of missingness depends on no observed and no unobserved quantities, and is therefore completely random. MAR allows missingness to depend on measured data and MNAR means that missingness may even relate to the unobserved data as well.

A classical result related to these concepts is that if one is using the complete cases (CC) only, i.e. those observations for which $C_i=1$, consistent estimates remain consistent if the missingness mechanism is MCAR. Under MAR, consistent estimates will typically not remain consistent using CC (though they often can be corrected\footnote{for example by using multiple imputation with a correctly specified imputation model}); but exceptions exist, for example when the probability of a value to be missing in the covariate $L$ does not depend on the outcome $Y$, and (non-causal) regression parameters are of interest; then consistent estimates remain consistent under a complete  case analysis \cite{Little:1992, Vach:1994}. Under MNAR it is more difficult to establish general conclusions, but it is commonly suggested that often consistent estimates won't remain consistent when using the complete cases only, again under exceptions -- such as when covariates in a regression model are missing only based on their own unobserved values, or in certain cases when missing values are replaced by missing data indicators \cite{Blake:2020}. Thus, a common practice is to omit missing data under MCAR, use methodological solutions (such as multiple imputation) under MAR and acknowledge limitations under MNAR, possibly accompanied by sensitivity analyses or particular models that make use of substantive assumptions about the missingness process \cite{Molenberghs:2009}.

Many authors have described the shortcomings of the above definitions: first, if multiple variables have missing values simultaneously, it is difficult to practically assess the plausibility of the MAR, MCAR, and MNAR assumptions \cite{Doretti:2018}. Second, whether a desired target parameter can be estimated, using the complete cases or differently, can not necessarily be determined with the M(N)AR framework alone \cite{Moreno:2018}. Lastly, it cannot be tested whether MAR, as defined above, holds in a given data set \cite{Mohan:2018}. To overcome these problems, an alternative graph-based framework has been proposed by Mohan and Pearl \cite{Mohan:2018}  because `` `the reasons for missingness' [...] is a causal, not a statistical concept'':
\begin{enumerate}[i)]
\item \textbf{Transparency.} This refers to the transparent classification of the missingness mechanism. The following steps are required:
\begin{enumerate}
\item[\textit{Step 1:}] Draw a canonical missingness DAG (m-DAG) which is a DAG that includes all missingness indicator variables $\textbf{C}=\{C_1,C_2,\ldots\}$. The m-DAG describes the assumptions about the data-generating process and the assumed causes of missingness.
\item[\textit{Step 2:}] Determine the conditional independencies implied by the m-DAG using graphical criteria (i.e. $d$-separation, see Section \ref{sec:assumptions_all} for details).
\item[\textit{Step 3:}] Define the graphical missingness mechanism. Let $\mathbf{L}=\{\mathbf{L}_1,\mathbf{L}_2\}$ where $\mathbf{L}_1$ contains the fully observed variables and $\mathbf{L}_2$ those that are partially observed and contain missing data. Data ($\mathbf{L},A,Y,\mathbf{C},U=\emptyset$) are said to be G-MCAR if $\mathbf{L}, A, Y \coprod \mathbf{C}$. G-MAR is fulfilled if the condition $\mathbf{L}_1, A, Y \coprod \mathbf{C} | \mathbf{L}_2$ is met. The data are defined to be G-MNAR if they are not G-MCAR or G-MAR. Mohan and Pearl \cite{Mohan:2018} provide the rules which can be used to determine the missingness mechanism in a given m-DAG\footnote{briefly, if there is no arrow between $\mathbf{C}$ and $\{\mathbf{L},A,Y\}$ the data are G-MCAR. If there is i) no arrow between any variable in $\mathbf{C}$ and any variable in $\mathbf{L}_2$ and ii) no path like $\mathbf{C} \leftarrow U \rightarrow \mathbf{L}_1$ the data are G-MAR.}. Note that G-MAR is a stronger requirement than MAR\footnote{essentially MAR refers to a \textit{set} of conditional independencies and the number of conditional independence assumptions that need to be verified is exponential in the number of missing variables, see Table 2 in Mohan and Pearl \cite{Mohan:2018} for an example. It may be difficult to practically argue for each of these dependency assumptions in a specific data example which is a weakness of the classic MAR definition; nevertheless, MAR is still the weakest known condition under which the missingness process can be ignored. Note that G-MAR and MAR are equivalent under the conditions of i) independent observations [as considered in this manuscript] and ii) conditional independence of the different missingness indicators.}.
\end{enumerate}
\item \textbf{Recoverability.} This refers to the task of establishing whether -- for a given m-DAG and given target parameter -- the latter can be estimated consistently from the observed data at all.
\begin{enumerate}
\item[\textit{Step 4:}] This is achieved by expressing the target parameter under the intervention of no missing data, e.g. as a counterfactual quantity that would have been observed if there had been no missing data -- and using the following conditions to evaluate if the desired target quantity can be expressed as a function of the observed (incomplete) data:
    \begin{enumerate}[a)]
    \item Consistency as defined in Section \ref{sec:assumptions6}. Since we intervene on $\mathbf{C}$ this equates to: if $C_Y=1$, then $Y^{C_Y=1}=Y$; and similarly if $C_A=1$, then $A^{C_A=1}=A$ and if $C_{\mathbf{L}_2}=1$, then $\mathbf{L}_2^{C_{\mathbf{L}_2}=\mathbf{1}}=\mathbf{L}_2$.\footnote{it can be argued \cite{Moreno:2018} that `intervening' on censoring due to death violates this assumption.}
    \item Positivity as defined in Section \ref{sec:assumptions5}. If we intervene on $C$, this is the requirement that $P(\mathbf{C}=\mathbf{1}|\mathbf{L_1=l_1}, \mathbf{L_2}^{C_{\mathbf{L}_2}=\mathbf{1}}=\mathbf{l_2}, A^{C_A=1}=a, Y^{C_Y=1}=y)>0$ for $\forall$ $\mathbf{l_1}, \mathbf{l_2}, a, y$ where we require $P(\mathbf{L_1=l_1}, \mathbf{L_2}^{C_{\mathbf{L}_2}=\mathbf{1}}=\mathbf{l_2}, A^{C_A=1}=a, Y^{C_Y=1}=y)>0$.
    \item Conditional independency assumptions as determined in Step 2. This is similar, yet not identical, to the weaker conditional exchangeability assumption from Section \ref{sec:assumptions1}.
    \item Factorization of the joint distribution of the data.
    \end{enumerate}
    If the target parameter is recoverable, use either existing results to obtain a suitable estimator (for example, using complete cases) or derive it manually. Briefly, under (G-)MCAR a complete case analysis is acceptable but not efficient; and under (G-)MAR existing methods (multiple imputation, weighting) may yield a consistent estimator if used correctly. For both (G-)MNAR and (G-)MAR estimands and estimators can be derived using i) the conditions a) - d); or ii) using existing results in the literature for specific m-DAG's \cite{Moreno:2018}; or iii) by utilizing specific theorems, for example Theorem 4 in Mohan and Pearl \cite{Mohan:2018}.
\end{enumerate}
\item \textbf{Testability.} This refers to the question whether it is possible to tell if any of the assumptions encoded in the m-DAG is incompatible with the observed data. MAR and MNAR is known to be untestable \cite{Allison:2001}. However, in any given data set it can be tested if G-MAR holds (Theorem 6 in Mohan and Pearl \cite{Mohan:2018}). Note that while G-MAR can be refuted, it can never be verified.
\begin{enumerate}
\item[\textit{Step 5:}] If G-MAR is assumed, test if the assumption meets the observed data.
\end{enumerate}
\end{enumerate}

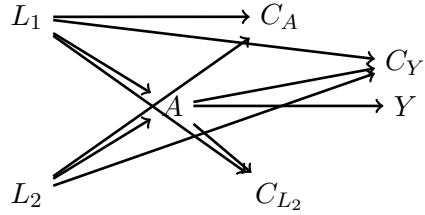
\begin{figure}[ht!]
\begin{center}
\begin{tikzpicture}
\node[text centered] (a) {$A$};
\node[left = 1.5 of a, text centered] (h1) {$ $};
\node[right = 1 of a, text centered] (h2) {$ $};
\node[right = 2 of a, text centered] (h3) {$ $};
\node[right = 2.5 of a, text centered] (y) {$Y$};
\node[above=0.75   of h1, text centered] (l1) {$L_1$};
\node[below=0.75   of h1, text centered] (l2) {$L_2$};
\node[above=0.75 of h2, text centered] (c1) {$C_A$};
\node[below=0.75 of h2, text centered] (c2) {$C_{L_2}$};
\node[above=0.01 of y,  text centered] (c3) {$C_Y$};
\draw [->, line width= 1] (a) --  (y);
\draw [->, line width= 1] (l1) --  (a);
\draw [->, line width= 1] (l2) --  (a);
\draw [->, line width= 1] (l1) --  (c1);
\draw [->, line width= 1] (l2) --  (c1);
\draw [->, line width= 1] (l1) --  (c2);
\draw [->, line width= 1] (a) --  (c2);
\draw [->, line width= 1] (l1) --  (c3);
\draw [->, line width= 1] (l2) --  (c3);
\draw [->, line width= 1] (a) --  (c3);
\end{tikzpicture}
\caption{Example of a m-DAG} \label{figure:mDAG}
\end{center}
\end{figure}

To better understand the above steps consider a simple example as depicted in the m-DAG in Figure \ref{figure:mDAG} (Step 1). The implied conditional independencies for the missingness indicators include\footnote{see Web Table 1 in Moreno-Betancur \cite{Moreno:2018}, Section 2.2 in Mohan and Pearl \cite{Mohan:2018} and Section \ref{sec:assumptions_all} on $d$-separation for more details on how the conditional independencies are derived.} $C_{L_2} \coprod L_2, Y |  L_1, A$; $C_A \coprod A, Y | L_1, L_2$; $C_Y \coprod Y | L_1, L_2, A$ (Step 2).

The data are G-MNAR because there exist arrows between $\mathbf{C}$ variables and partially observed variables, for example $L_2 \rightarrow C_A$ (Step 3). Thus, existing results for classic MAR with respect to complete case analyses or multiple imputation (or others) can not necessarily be well justified. In the regression context, we are interested in the conditional expectation of $Y$ given $A$ and $\mathbf{L}$ (and how this is defined in terms of regression coefficients). To establish whether the full conditional distribution can be recovered we can use the conditions a) - d) from Step 4 to derive the following:

\begin{eqnarray*}
P(Y^{\mathbf{C}=\mathbf{1}}|A^{\mathbf{C}=\mathbf{1}},L_2^{\mathbf{C}=\mathbf{1}},L_1) &=& P(Y^{\mathbf{C}=\mathbf{1}}|A^{\mathbf{C}=\mathbf{1}},L_2^{\mathbf{C}=\mathbf{1}},L_1, \mathbf{C}=1) \\
&=& P(Y|A,L_2,L_1,\mathbf{C}=\mathbf{1})
\end{eqnarray*}

The left-hand side states the target distribution under the intervention that there is no missing data, i.e. that $\mathbf{C}$ is set to $\mathbf{1}$ for everyone. The first equality follows from the conditional independence assumptions; the second from the assumption of consistency. Positivity is required because we want the conditional distribution to be well-defined for all covariate strata. A factorization of the distribution was not needed in the given example. The established result shows that a complete case analysis with regression is valid and the missing-data mechanism allows us to estimate the desired target parameter. Instead of deriving the above results, it would also have been possible to consult Moreno-Betancur et al. \cite{Moreno:2018} to verify that for the given m-DAG recoverability is achieved by a complete case analysis. Step 5 is not needed here because G-MAR is not assumed.

In summary, if there is no missing data, then there is no problem. If there is missing data, and if the missingness mechanism is indeed MCAR, fitting a regression model on the complete cases is valid\footnote{however, at the expense of loss of information \cite{Little:1992}}. Under MAR established results regarding consistent estimation of regression parameters can be used \cite{Little:1992}, for example using multiple imputation or sometimes complete case analyses; but if multiple variables are missing, MAR can neither be justified well nor tested in the data. Thus, it may be better to state one's assumption about the missingness process in a m-DAG. Following the graphical framework of Mohan and Pearl \cite{Mohan:2018}, namely transparency, recoverability and testability, one can establish whether the the desired target quantity can be estimated or not; an if yes, how. Under (G-)MNAR the graph-based framework offers to date the best approach to decide whether the question at hand can be answered or not.

Given the above considerations it follows that using regression models on incomplete data can lead to bias in causal effect estimation, and whether this is the case, and how it can be addressed, can be determined by making structural assumptions about the data-generating process.

\subsection{Assumption 12: no relevant measurement error}\label{sec:assumptions12}
Measurement error (ME), also known as mis-classification for categorical variables\footnote{non-compliance as described under assumption 9 can also be seen as measurement error because the measured treatment $A$ deviates from the assigned treatment $\tilde{A}$.}, will often lead to bias when estimating the causal effect of $A$ on $Y$. Estimates of regression coefficients can either be attenuated or strengthened by measurement error \cite{Carroll:2006}. Similar to the former section a simple summary about whether bias exists or not can be determined by evaluating back-door paths in DAGs and by taking into account the process that leads to the measurement error.

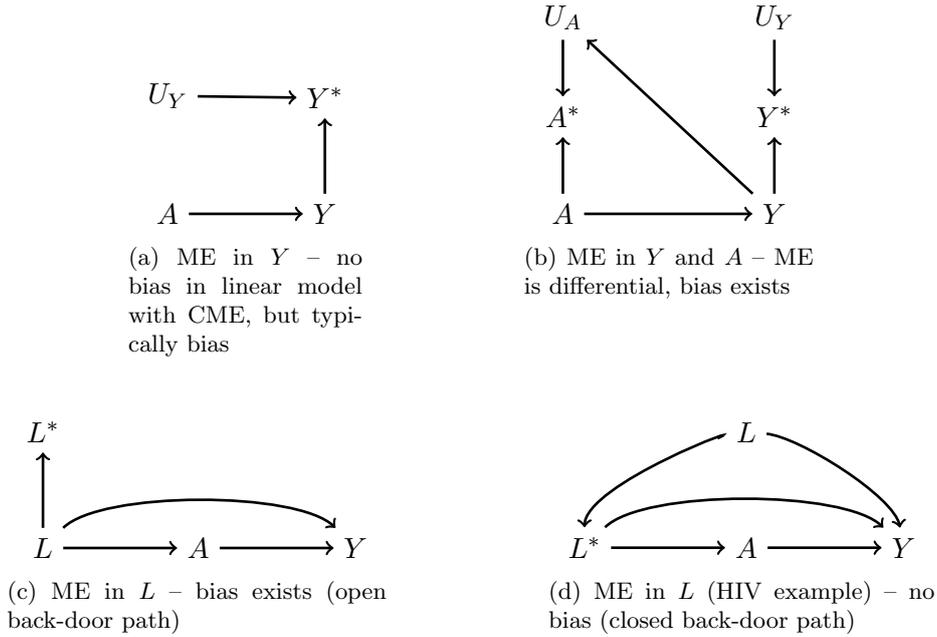
\begin{figure}[ht!]
\begin{center}
\subfloat[ME in $Y$ -- no bias in linear model with CME, but typically bias]{\label{figure:DAG5a}
\begin{tikzpicture}
\node[text centered] (a) {$A$};
\node[right=1.5 of a, text centered] (y) {$Y$};
\node[above = 1 of y, text centered] (ys) {$Y^\ast$};
\node[above = 1 of a, text centered] (uy) {$U_Y$};
\draw[->, line width= 1] (uy) -- node[above,font=\footnotesize]{}  (ys);
\draw[->, line width= 1] (y) -- node[above,font=\footnotesize]{}  (ys);
\draw[->, line width= 1] (a) -- node[above,font=\footnotesize]{}  (y);
\end{tikzpicture}
}\hspace*{2cm}
\subfloat[ME in $Y$ and $A$ -- ME is differential, bias exists]{\label{figure:DAG5b}
\begin{tikzpicture}
\node[text centered] (ua) {$U_A$};
\node[right = 2 of ua, text centered] (uy) {$U_Y$};
\node[below=0.75 of ua, text centered] (as) {$A^\ast$};
\node[below=0.75 of uy, text centered] (ys) {$Y^\ast$};
\node[below=0.75 of as, text centered] (a) {$A$};
\node[below=0.75 of ys, text centered] (y) {$Y$};
\draw[->, line width= 1] (ua) -- node[above,font=\footnotesize]{}  (as);
\draw[->, line width= 1] (uy) -- node[above,font=\footnotesize]{}  (ys);
\draw[->, line width= 1] (a) -- node[above,font=\footnotesize]{}  (as);
\draw[->, line width= 1] (y) -- node[above,font=\footnotesize]{}  (ys);
\draw[->, line width= 1] (a) -- node[above,font=\footnotesize]{}  (y);
\draw[->, line width= 1] (y) -- (ua);
\end{tikzpicture}
}\\[0.3cm]
\subfloat[ME in $L$ -- bias exists (open back-door path)]{\label{figure:DAG5c}
\begin{tikzpicture}
\node[text centered] (ls) {$L^\ast$};
\node[below=1 of ls, text centered] (l) {$L$};
\node[right=1.5 of l, text centered] (a) {$A$};
\node[right=1.5 of a, text centered] (y) {$Y$};
\draw[->, line width= 1] (l) -- node[above,font=\footnotesize]{}  (ls);
\draw[->, line width= 1] (l) -- node[above,font=\footnotesize]{}  (a);
\draw[->, line width= 1] (a) -- node[above,font=\footnotesize]{}  (y);
\draw[->, line width=1] (l) to [out=45,in=135, looseness=0.5] (y);
\end{tikzpicture}
}\hspace*{2cm}
\subfloat[ME in $L$ (HIV example) -- no bias (closed back-door path)]{\label{figure:DAG5d}
\begin{tikzpicture}
\node[text centered] (t) {$L^{\ast}$};
\node[right = 1.5 of t, text centered] (m) {$A$};
\node[right=1.5 of m, text centered] (y) {$Y$};
\node[above=1 of m, text centered] (l) {$L$};
\draw[->, line width= 1] (t) -- node[above,font=\footnotesize]{}  (m);
\draw[->, line width=1] (t) to  [out=45,in=135, looseness=0.5] node[below,font=\footnotesize]{} (y);
\draw[->, line width=1] (m) to (y);
\draw[->, line width=1] (l) to [out=-5,in=100, looseness=0.5] (y);
\draw[->, line width=1] (1.75,1.45) to [out=25,in=90, looseness=0.7] (t);
\end{tikzpicture}
}
\caption{Examples for different measurement error mechanisms} \label{figure:DAGs-measurement error}
\end{center}
\end{figure}

Figure \ref{figure:DAG5a} shows a simple example of measurement error: there are no confounders, $A$ is measured accurately, but the measured outcome ($Y^{\ast}$) does not correspond to the actual outcome. One may add an unmeasured variable $U_Y$ to the DAG to describe any factors (other than the actual outcome) that influence the measured $Y$. In general there is no guarantee that, for example, the associational difference $E(Y|A^{\ast}=1)-E(Y|A^{\ast}=0)$ corresponds to the causal difference $E(Y^{A=1}) - E(Y^{A=0})$. It is however possible to ignore measurement error in the outcome under specific assumptions: for instance, if $Y|A$ follows a normal distribution (i.e. a linear model is assumed) and one has classical measurement error (CME), i.e.  $Y^{\ast}_i = Y_i + \upsilon_i$, $\upsilon_i \sim N(0,\sigma^2_{\upsilon_i})$. Since $Y^{\ast}_i = Y_i + \upsilon_i$, we have $Y^{\ast}_i = Y_i + \beta_0 + \beta_1 A_i + \epsilon_i + \upsilon_i$, $\epsilon_i \sim N(0,\sigma^2_{\epsilon_i})$, and consistent estimates remain consistent, at least if $\epsilon$ and $\upsilon$ are independent.

It is however unusual that such (or similar) simple assumptions would hold in many realistic settings. The measurement error may also not be additive and caused by various measured and unmeasured variables. While ignoring measurement error often leads to bias with respect to the causal target quantity, it makes still sense to evaluate the structure of the measurement error as for certain types of error, regression estimates can be corrected if data or knowledge about the measurement error process is existent. Figure \ref{figure:DAG5b} gives an example of independent but differential measurement error. The ME is called independent because $U_A$ and $U_Y$ (which relate to the ME process of both $Y$ and $A$) are independent given that the path between them is blocked by the collider $Y^{\ast}$; the ME is however differential because $U_A$ depends on the outcome $Y$ \footnote{if $U_Y$ depends on $A$ the ME is differential too.}. Depending on whether the ME is dependent or independent, and differential or non-differential, and depending which data is measured, it may be possible to correct for the bias of the regression coefficients; see Buonaccorsi \cite{Buonaccorsi:2010} and Carroll et al. \cite{Carroll:2006} for details.

Confounders or colliders may be mis-measured too. Conditioning on colliders with measurement error still leads to collider bias because the mismeasured variable is still a cause\footnote{that is, a descendant in a DAG.} of both $A$ and $Y$ \cite{Hernan:2020}. Whether ME in a confounder causes bias with respect to effect estimation depends on the respective situation. In Figure \ref{figure:DAG5c} a mis-measured confounder causes bias because it does not block the back-door path $A \leftarrow L \rightarrow Y$. But in Figure \ref{figure:DAG5d} there is no problem in using the mis-measured variable because all back-door paths are blocked by conditioning on $L^{\ast}$. An example for a situation like in \ref{figure:DAG5d} is when HIV-positive patients are assigned antiretroviral treatment ($A$) based on their measured laboratory values (CD4 count, $L^{\ast}$); while the true underlying disease severity markers ($L$) influence both the measured markers and the outcome (say death), the measured values will still be a strong marker to determine mortality and conditioning on them will block the open back-door path.

In summary, measurement error will often lead to biased regression coefficients, except in either very simple situations or situations where mis-measured variables act as a confounder that can block a back-door path. If the structure (in/dependent; non-/differential) of the measurement error is known, it may be possible to correct bias in regression coefficients.

\subsection{Summary of all assumptions}\label{sec:assumptions_all}
The above introduced assumptions intersect with each other and it is possible to distill them into a smaller set of assumptions.  To do so, it is useful to understand the concept of $d$-separation. A set of variables $\mathcal{\tilde{S}}$ is said to block, or \textit{d-separate}, a path $p$ if either
\begin{enumerate}[i)]
\item $p$ contains a chain\footnote{i.e., a sub-path of $p$} $A \rightarrow L \rightarrow Y$ or a fork $A \leftarrow L \rightarrow Y$ such that $L$ is in $\mathcal{\tilde{S}}$ or
\item $p$ contains an inverted fork $A \rightarrow L \leftarrow Y$ such that $L$ is not in $\mathcal{\tilde{S}}$, and no descendant of $L$ either.
\end{enumerate}
If $\mathcal{\tilde{S}}$ blocks all paths from $A$ to $Y$, then $A$ and $Y$ are independent conditional on $\mathcal{\tilde{S}}$. The $d$-separation definition implies that $A$ and $Y$ are conditionally independent given $\mathcal{S}$ if back-door paths between $A$ and $Y$ are blocked, which can be achieved by conditioning on a confounder or non-conditioning on a collider. At this point it is useful to also recall the \textit{back-door criterion}\cite{Pearl:2009}, which states that a set $\mathcal{S}$ is sufficient for causal effect identification if i) the elements of $\mathcal{S}$ block all back-door paths from $A$ to $Y$ (i.e. $d$-separate $A$ and $Y$ and thus guarantee conditional exchangeability) and ii) $\mathcal{S}$ does not contain a descendant of $A$. This theorem echoes much of the assumptions 1, 2, 3, 4, 10 and 12 and thus the first crucial assumption can be summarized as:
\begin{flalign*}
\text{(AS1)}: & \,\text{the set of variables $\mathcal{S}=\mathbf{L}$ satisfies the back-door criterion}\,.&
\end{flalign*}

Assumption 4 was the assumption of positivity\footnote{it may be possible to relax this assumption under additional parametric assumptions, see Greenland (2017) \cite{Greenland:2017} for examples}:
\begin{flalign*}
\text{(AS2)}:\quad\quad&   P(A=a|\mathbf{L}=\mathbf{l})>0 \quad \text{for} \quad \forall \mathbf{l} \quad \text{with} \quad P(\mathbf{L}=\mathbf{l})\neq 0.&\\
\end{flalign*}

Assumptions 6 and 7 relate to consistency and are, as discussed above, sometimes stated as a theorem rather than an assumption \cite{Pearl:2010b}:
\begin{flalign*}
\text{(AS3)}:\quad\quad& \text{If} \, A=a, \,\text{then}\, Y^a=Y \quad \text{for}\,\forall a\,.&
\end{flalign*}
Assumptions 8 and 9 directly target the definition of the target quantity (and aren't necessary assumptions for causal inference in general). We call them (AS4) and (AS5) here. For purpose of illustration of the example below, we may define them in a somewhat over-simplified\footnote{see Pearl (2009) \cite{Pearl:2009} for a general definition of collapsibility and vander Weele (2012) \cite{vanderWeele:2012} for effect modification} way as:

\begin{flalign*}
\text{(AS4)}:\quad\quad&   \text{collapsibility; \quad for example, that} \quad \beta_1 = \beta_1^{\ast} \quad \text{for} &\\ & \quad E(Y|A=a, L=l) = \beta_0 + \beta_1 a + \beta_2 l \quad \text{and} &\\
& \quad E(Y|A=a) = \beta_0 + \beta_1^{\ast} a &\\
&\text{even if L is not a confounder and is related only to Y and not A} \cite{Greenland:1999, Sjolander:2016, Pang:2016}\,.&\\
\text{(AS5)}:\quad\quad&   \text{no effect modification; \quad for example, that} &\\
&E(Y^a|L=l_1) = E(Y^a|L=l_0)\,.&
\end{flalign*}

To understand how these these assumptions are needed in a specific example, assume that the ATE is the causal quantity of interest, and that all relationships of interest are linear and a linear model is the regression model of interest, i.e. $E(Y|A,L)=\beta_0 + \beta_1 A + \beta_2 AL + \beta_3 L$. The average treatment effect can then be identified using the regression parameter $\beta_1$ under assumptions (AS1)-(AS5) because

\begin{eqnarray*}
\text{ATE} &=& E(Y^1-Y^0) = E(Y^1) - E(Y^0) \\
           &\stackrel{(AS1)}{=}& E(Y^1|A=1,L=l) - E(Y^0|A=0,L=l)\\
           &\stackrel{(AS3)}{=}& E(Y|A=1, L=l) - E(Y|A=0, L=l)\\
           &\stackrel{(AS2)}{=}& \beta_0 + \beta_1  + \beta_2 l+ \beta_3 l - \beta_0 - \beta_3 l = \beta_1  + \beta_2 l \\
           &\stackrel{(AS4, AS5)}{=}& \beta_1\,.
\end{eqnarray*}

The equalities follow directly from the above definitions. Positivity is required because the conditional expectation $E(Y|A=a,L=l)$ may not be well-defined under the model $\beta_0 + \beta_1 A + \beta_2 AL + \beta_3 L$ if the model is not saturated and the stratum $A=a, L=l$ is empty or sparse because of $P(A=a|L=l)\approx 0$. The last equality only holds if for all covariate strata $L=l$ the same results are obtained; which implies no effect modification and collapsibility. If there is no effect modification, there is still $\beta_1 \neq  \beta_1  + \beta_2 l$ for any $l$ under non-collapsibility.

Thus, any consistent estimator $\hat{\beta}_1$ of $\beta_1$ consistently estimates the ATE\footnote{this implies that using estimates conditional on a prior model selection step, or using shrinkage estimation techniques, would be invalid \cite{Chatfield:1995, Leeb:2005, Schomaker:2012}.}. If there is missing data, this may not necessarily be the case and thus a last assumption is:
\begin{flalign*}
\text{(AS6)}:\quad\quad&    \text{The process generating the missing data allows the use of regression, }&\\
& \text{for example based on the complete cases.} &
\end{flalign*}

As explained in Section \ref{sec:assumptions11}, whether this is the case, is complicated and needs to be determined on a case-by-case basis.

\subsection{The sample selection process: are further assumptions required?}\label{sec:assumptions_sample}
Causal effects are implicitly defined with respect to a population of interest\footnote{that is, the actual data may consist of $n$ i.i.d. copies from $\mathcal{D}=(L,A,Y) \sim P_0$, where $P_0$ is the distribution of $\mathcal{D}$ with respect to a target population; for example, patients that are eligible for intervention assignment based on certain demographics.}. However, the actual sample may not necessarily be a random draw of the target population. As highlighted in the literature \cite{Infante:2018}, this may have the following reasons:
\begin{enumerate}[i)]
\item selection into the sample \textit{before} intervention assignment, for example due to study inclusion/exclusion criteria or sampling designs/mechanisms.
\item selection into the study based on mechanisms \textit{after} intervention assignment, for example due to non-response, drop-out, censoring, subset-analyses or study design (e.g., case-control studies).
\end{enumerate}

\begin{figure}[ht!]
\begin{center}
\subfloat[adjustment for confounder $L_2$ is sufficient]{\label{figure:DAG6a}
\begin{tikzpicture}
\node[text centered] (a) {$A$};
\node[right = 1 of a, text centered] (y) {$Y$};
\node[draw,left  = 1 of a, text centered] (s) {$S$};
\node[left  = 1 of s, text centered] (help) {};
\node[above  = 1 of help, text centered] (l2) {$L_2$};
\node[left  = 1 of l2, text centered] (l1) {$L_1$};
\draw[->, line width= 1] (a) -- (y);
\draw[->, line width= 1] (l2) -- (y);
\draw[->, line width= 1] (l2) -- (a);
\draw[->, line width= 1] (l1) -- (a);
\draw[->, line width= 1] (l1) -- (s);
\end{tikzpicture}
}\hspace*{2cm}
\subfloat[adjustment for confounder $L_2$ is sufficient]{\label{figure:DAG6b}
\begin{tikzpicture}
\node[text centered] (a) {$A$};
\node[right = 1 of a, text centered] (y) {$Y$};
\node[draw,left  = 1 of a, text centered] (s) {$S$};
\node[left  = 1 of s, text centered] (help) {};
\node[above  = 1 of help, text centered] (l2) {$L_2$};
\node[left  = 1 of l2, text centered] (l1) {$L_1$};
\draw[->, line width= 1] (a) -- (y);
\draw[->, line width= 1] (l2) -- (y);
\draw[->, line width= 1] (l2) -- (a);
\draw[->, line width= 1] (l1) -- (y);
\draw[->, line width= 1] (l1) -- (s);
\end{tikzpicture}
}\\[0.3cm]
\subfloat[adjustment for either $L_1$ or $L_2$, or both, is needed (M-Bias)]{\label{figure:DAG6c}
\begin{tikzpicture}
\node[text centered] (a) {$A$};
\node[right = 1 of a, text centered] (y) {$Y$};
\node[draw,left  = 1 of a, text centered] (s) {$S$};
\node[left  = 1 of s, text centered] (help) {};
\node[above  = 1 of help, text centered] (l2) {$L_2$};
\node[left  = 1 of l2, text centered] (l1) {$L_1$};
\draw[->, line width= 1] (a) -- (y);
\draw[->, line width= 1] (l2) -- (y);
\draw[->, line width= 1] (l1) -- (l2);
\draw[->, line width= 1] (l1) -- (a);
\draw[->, line width= 1] (l1) -- (s);
\draw[->, line width= 1] (l2) -- (s);
\end{tikzpicture}
}\hspace*{2cm}
\subfloat[with dashed arrow, $L_2$ needs to be adjusted for]{\label{figure:DAG6d}
\begin{tikzpicture}
\node[text centered] (a) {$A$};
\node[right = 1 of a, text centered] (y) {$Y$};
\node[draw,left  = 1 of a, text centered] (s2) {$S_2$};
\node[draw,left  = 1 of s2, text centered] (s1) {$S_1$};
\node[left  = 1 of s1, text centered] (help) {};
\node[above  = 1 of s1, text centered] (l2) {$L_2$};
\node[left  = 1 of l2, text centered] (l1) {$L_1$};
\draw[->, line width= 1] (a) -- (y);
\draw[->, line width= 1, dashed] (l2) -- (y);
\draw[->, line width= 1] (l2) -- (a);
\draw[->, line width= 1] (l1) -- (a);
\draw[->, line width= 1] (l1) -- (s1);
\draw[->, line width= 1] (l2) -- (s2);
\draw[->, line width= 1] (l2) -- (s1);
\draw[->, line width= 1] (s1) -- (s2);
\end{tikzpicture}
}
\caption{Examples for different sample selection mechanisms. Here, sample selection is determined before intervention assignment} \label{figure:DAGs-sample_selection}
\end{center}
\end{figure}
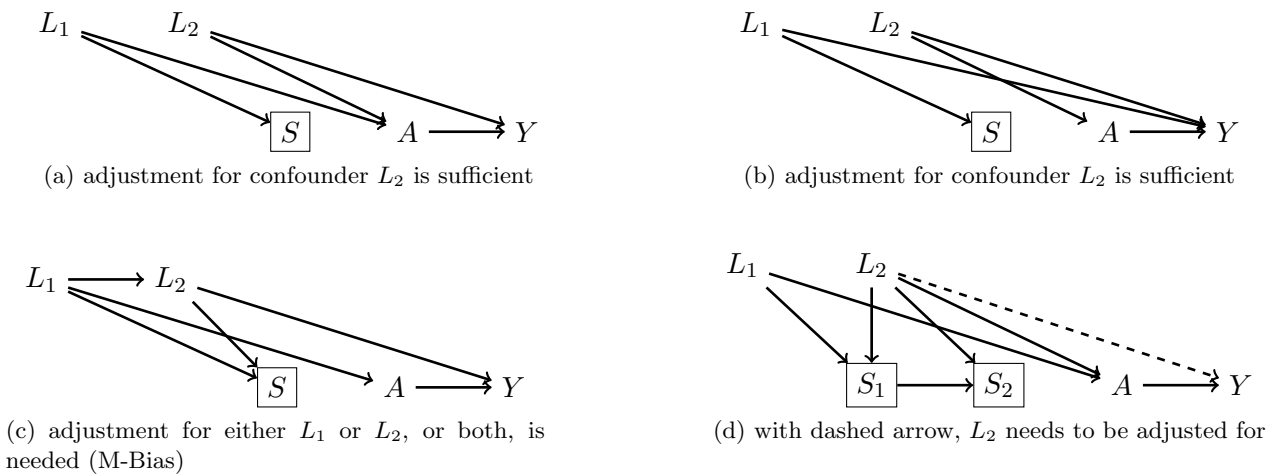

The situations listed under ii) have been dealt with above already: for example, in the section on colliders (Figure  \ref{figure:DAG3c}), mediators (e.g., Figure  \ref{figure:DAG_med_c}), considerations of per protocol selection bias (i.e., conditioning on $A=\tilde{A}=1$)  and validity of complete case analyses (Section \ref{sec:assumptions10}). Whether regression coefficients can thus be used for causal effect estimation under a particular sample selection process can therefore be evaluated using DAGs and the back-door theorem: selection indicators $S$\footnote{in previous examples, the variable $C$ was used for censoring and missing data mechanisms and would correspond to the variable $S$ in this section} are simply added to the DAG, conditioned on, and the relevant back-door paths from $A$ to $Y$ are evaluated. Similarly, one can evaluate the situations described under i), see Figure \ref{figure:DAGs-sample_selection} for examples.

In all 4 Figures, $L_1$ and $L_2$ \textit{precede} study selection and $S$ ($S_1, S_2$) is being conditioned on. In Figure \ref{figure:DAG6a}, sample inclusion is based on $L_1$ (e.g., region) but this does not affect back-door paths from $A$ to $Y$ and adjusting for the confounder $L_2$ is sufficient. The same applies to the situation in Figure \ref{figure:DAG6b}. Figure \ref{figure:DAG6c} shows a typical situation where sample selection equates to conditioning on a collider. The path $L_1 \rightarrow S \leftarrow L_2$ is opened; the relevant back-door path $A \leftarrow L_1 \rightarrow \fbox{S} \leftarrow L_2 \rightarrow Y$ needs to be closed by adjusting for either $L_1$ or $L_2$ (or both). Without adjustment, M-bias\footnote{this is because the path $A \leftarrow L_1 \rightarrow S \leftarrow L_2 \rightarrow Y$ looks like an ``M'' if the DAG is drawn differently, with no consideration of time-ordering} will occur. Figure \ref{figure:DAG6d} shows a typical situation of a two-phase sampling process. For example, $S_1$ may be an indicator for being eligible for a study and being contacted by the study team, whereas $S_2$ may be an indicator to respond to an invitation to participate. For the given sample selection process, no adjustment of $L_1$ or $L_2$ is needed if there was actually no dashed arrow. Otherwise, adjustment of $L_2$ would be required to close the back-door paths that were opened by conditioning on $S_1$ and $S_2$ respectively.

To answer the question posed in this section: to determine appropriate adjustment sets $\mathcal{\tilde{S}}$ for regression, assumption (AS1) related to the back-door theorem as well as the considerations from the missing data Section \ref{sec:assumptions10} remain valid if selection indicators are included in the DAG (after the intervention variable, selection indicators may, for example, correspond to missing data or censoring indicators). For the somewhat limited purpose of this paper, it is therefore not needed to introduce further assumptions. However, it is important to highlight that the more general question of identifying and estimating causal effects for target populations that are different from the study population may require refinement of assumptions. The generalizability and transportability literature discusses if and how causal effects can be recovered for given sample selection processes \cite{Degtiar:2021}. It turns out that a more general selection back-door criterion, or an exchangeability assumption regarding the selection indicator(s), is required in this case \cite{Pearl:2015, Bareinboim:2016} and estimation is best done with standardization and weighting-based approaches.

In summary, adding selection indicators to the structural model (DAG) aids the verification of assumptions (AS1) and (AS6) from Section \ref{sec:assumptions_all}. The more ambitious task of transporting causal effect estimates to target populations require additional results and methods that are typically unrelated to regression coefficient estimation and interpretation, and are therefore not dealt with in this paper.

\section{Observational Data and Randomized Studies}\label{sec:obsdata_and_expdata}

Some of the assumptions outlined above can be met by design. It is useful to distinguish between randomized studies (i.e., studies or experiments where intervention assignment is under control of an experimenter) and observational studies (where intervention assignment is not necessarily under control of someone).

For observational studies none of the assumptions can be expected to be met by design; and it may thus be advisable to check them. However, for randomized studies the study design guarantees that \textit{some} assumptions are met. Table \ref{tab:RS_vs_OS} gives a summary on the design-assumptions relationship.

\begin{table}[ht!]
\caption{The checking of assumptions depending on study design}\label{tab:RS_vs_OS}
\begin{center}
\fbox{
\begin{tabular}{p{0.145\textwidth}|p{0.34\textwidth}|p{0.2\textwidth}|p{0.16\textwidth}}
& \multicolumn{1}{c|}{Randomized Studies} & \multicolumn{1}{c|}{Observational Studies} & \multicolumn{1}{c}{where to check} \\
\hline
Assumption 1  &$\checkmark$ achieved by design & \textbf{!} needs to be verified & causal model\\
Assumption 2  &($\checkmark$) typically not relevant& \textbf{!} needs to be verified & stat. model\\
Assumption 3  &(!) needs to be verified sometimes & \textbf{!} needs to be verified & causal model\\
Assumption 4  &\textbf{!} needs to be verified&\textbf{!} needs to be verified & causal model\\
Assumption 5  &$\checkmark$ achieved by design & \textbf{!} needs to be verified & stat. model\\
Assumption 6  &($\checkmark$) met, if well-designed& \textbf{!} needs to be verified & causal/concept.\\
Assumption 7  &($\checkmark$) often not a problem & \textbf{!} needs to be verified &  causal/concept.\\
Assumption 8  & ($\checkmark$) typically not relevant by design & \textbf{!} needs to be verified & stat. model\\
Assumption 9  &\textbf{!} needs to be verified&\textbf{!} needs to be verified & stat. model\\
Assumption 10 &\textbf{!} needs to be verified&\textbf{!} similar to ass. 12 & causal/concept.\\
Assumption 11 &\textbf{!} needs to be verified&\textbf{!} needs to be verified & causal/statist.\\
Assumption 12 &\textbf{!} needs to be verified&\textbf{!} needs to be verified & causal model\\
\end{tabular}
}
\end{center}
\end{table}

First, a (completely) randomized experiment requires by definition that $0 < P(A_i=1) < 1$. Thus, the assumption of positivity is always met. Another implication is that the treatment assignment process does (trivially) not depend on potential outcomes; that is, $P(A_i=1|Y^a) = P(A_i=1) \quad \forall A=a$. This requirement can also be framed as $P(Y^a=1|A=1) = P(Y^a=1|A=0) = P(Y^a=1)$ which makes it immediately clear that exchangeability ($Y^{{a}} \coprod {A}$) is achieved by design. This implies that randomized experiments don't face the problem of unmeasured confounders (assumption 1). Stratified or paired randomized experiments only guarantee conditional exchangeability, not exchangeability, which is however sufficient for assumption 1 to be met \cite{Imbens:2015}.

Correct model specification (assumption 2) is usually not a problem for completely randomized studies, either because covariate inclusion is not necessary to identify the causal effect or because hypothesis tests are often still valid under model mis-specification \cite{Rosenblum:2009}. Of course, if $A$ is continuous and the dose-response relationship between $A$ and $Y$ is of interest, model mis-specification of this relationship may still be a concern.

In many randomized experiments, variables which are measured after the outcome (endpoint) are not included in the analysis. Thus, the threat of conditioning on a collider (assumption 3) is often limited. Nevertheless, in some situations collider bias could still be a problem, for example in time-to-event analyses where a particular outcome is evaluated on a subset of survivors \cite{Schomaker:2020}. Mediators are post-intervention variables which are not included in analyses in some fields \cite{Imbens:2019}. However, reflecting on mediation (and assumption 4) may still be desirable in many areas as variables measured between intervention assignment and the outcome may be captured and potentially used in analyses.

A well-designed randomized experiment, with a clear study protocol, should typically not suffer from violations of the consistency assumption. However, there might be situations in which violations could occur: for example, if a surgical procedure was performed differently by different surgeons, then the consistency assumption may be violated and there may be multiple versions of treatment -- all with different effects on the outcome. Many considerations regarding the consistency assumption also refer to the SUTVA assumption; however, interference between study subjects may be a particular concern and can be problematic in randomized studies: if interference itself is not the subject of interest, applying standard regression approaches to estimate common treatment effects will yield invalid inference \cite{Luo:2012, Rosenbaum:2007}.

As opposed to observational studies, effect modification is no concern at all for completely randomized experiments. As noted above, a completely randomized experiment does not only achieve conditional exchangeability by design, but even exchangeability. Thus, $P(Y^a=1|A=1,L=l) = P(Y^a=1|A=0,L=l) = P(Y^a=1)$ for a binary treatment and $\forall L=l$. In stratified or paired randomized experiments effect modification could in principle prohibit marginal effect estimation with regression; however, in this case there is only a limited set of strata one would condition on; and thus the change to conditional target quantities would be less of a concern as in observational studies.

Collapsibility (assumption 9) is a relevant consideration for randomized studies. Using non-collapsible effect measures, for example the conditional odds ratio estimated with logistic regression, can lead to bias with respect to marginal target quantities. It follows that the gain in precision that may come with inclusion of covariates $\mathbf{L}$ in regression models comes at the cost of bias in randomized studies.

Of course, compliance, missing data (drop out) and measurement error are relevant topics for randomized experiments  too, and assumptions 10-12 need to be checked. A more comprehensive discussion on assumptions in randomized trials follows in Section \ref{sec:implications_randomized_trials}.

Independent of study design, it is important to reflect if and how assumptions can be checked. Table \ref{tab:RS_vs_OS} gives a summary on when the causal model (e.g., the DAG) can be used for it, when the statistical model can be used (and assumptions actually tested in the data set) or whether only conceptual considerations are useful.

A prominent view in the literature ist that assumptions on whether variables are confounders, colliders and mediators (i.e., assumptions 1,3,4,12) need to be checked in the causal model and cannot be verified empirically in the data -- with the exception that the variables included in the statistical (regression) model should be those determined through the causal model and back-door theorem. This paper follows the perspective that the absence of arrows in a DAG is purely knowledge-based \cite{Petersen:2014} and the encoded structural mechanisms unambiguously define the role of variables. An alternative or complementary view would be that additional knowledge should be gained through causal learning algorithms, which can be applied in some settings to disentangle structures and thus implicitly discover roles of variables. This angle goes however beyond the scope of this paper and the reader is referred to the literature \cite{Eberhard:2017, Glymour:2019}.

Whether the positivity assumption is possibly violated can be partly checked in the data: if the treatment is binary (or categorical), one could estimate the propensity scores $P(A=1|\mathbf{L=l})$ and $P(A=0|\mathbf{L=l})$  in the data and summarize the estimated probabilities in kernel density plots \cite{Luque:2018}. Small estimated probabilities indicate practical positivity violations. However, structural positivity violations where treatment assignment within particular strata is prohibited or infeasible, can not be discovered with this approach. Moreover, if there are no measured data for certain covariate strata, also other practical positivity violations may remain undiscovered.

Bias due to incorrect model specification can not be clearly verified in the data; however, using flexible modeling strategies such as penalized splines within additive regression models, may help to reduce the risk of model misspecification. This strategy may also help to detect (non-linear) interactions, i.e. check assumption 8. Nevertheless, discovering highly non-linear structures and complex (higher order) interactions remains a challenge -- and applying data adaptive methods to tackle this issue and estimate regression parameters more flexibly is incorrect as the uncertainty associated with estimating tuning parameters is not adequately reflected in such inference procedures \cite{Chatfield:1995, Hjort:2003, Schomaker:2012}.

In theory, considerations with respect to both the consistency and interference assumption can be reflected in the causal model: different versions of treatment can be visualized in a DAG \cite{Pearl:2010b} and networks of interactions among units as well \cite{Ogburn:2014}. However, practically, these assumptions may be checked and discussed on a more conceptual level. For example, \textit{retrospectively} one may ask: was it likely that different surgeons performed a surgery differently and would this affect estimation and interpretation of the causal effect? Would it make sense to speak more to the surgeons and discuss the consistency assumption? Or: was it likely that study participants interacted in a way that could violate the interference assumption? Would violations be rare and could they therefore be possibly ignored? As such, the appropriateness of assumptions 6 and 7 may in many circumstances not be fully clear and could be checked through discussions and interactions, rather than integration into the structural model.

Whether there is compliance with treatment assignment is typically knowledge-based, reflected in the structural model and considered during estimation. However, this assumption cannot be ``tested'' in the data.

Knowledge about collapsible and non-collapsible effect measures may affect the estimand choice and ultimately the choice of the statistical model. For example, if the scientific question translates into a marginal estimand, which is most often the case, the knowledge that the risk ratio is a collapsible effect measure whereas the odds ratio is not, would guide the analyst to fitting a Poisson regression model, rather than a logistic regression model.

Lastly, how does one verify whether the missingness process prohibits causal effect estimation with regression parameters? In simple cases, this can actually be checked in the data: suppose only one variable is missing and based on prior knowledge it can be plausibly defended that the data are not MNAR, then a regression of the binary missingness indicator on the covariates can support either a MAR or MCAR assumption and therefore the analysis choice, for example a complete case analysis. As highlighted in Section \ref{sec:assumptions11}, in more complex situations it is beneficial to encode assumptions about the missingness process in a DAG. This equates to verifying assumption 11 in the causal model; however, whether a G-MAR or G-MCAR assumption is violated can in principle be tested in the data, see Section 4.2 in Mohan and Pearl \cite{Mohan:2018} for details. Practical guidelines on how to utilize this in sophisticated and realistic analyses, remains to be investigated and demonstrated.

\section{Illustration}\label{sec:illustration}
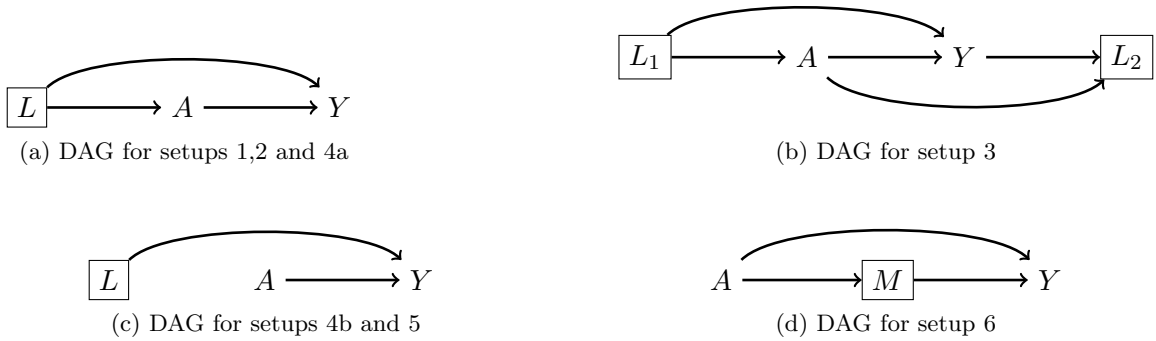
\begin{figure}[ht!]
\begin{center}
\subfloat[DAG for setups 1,2 and 4a]{\label{figure:DAG_sim_a}
\begin{tikzpicture}
\node[draw,text centered] (l) {$L$};
\node[right = 1.5 of l, text centered] (a) {$A$};
\node[right = 1.5 of a, text centered] (y) {$Y$};
\draw [->, line width= 1] (a) -- node[above,font=\footnotesize]{}  (y);
\draw [->, line width= 1] (l) -- node[above,font=\footnotesize]{}  (a);
\draw[->, line width=1] (l) to  [out=45,in=135, looseness=0.5] node[below,font=\footnotesize]{} (y);
\end{tikzpicture}
}\hspace*{3cm}
\subfloat[DAG for setup 3]{\label{figure:DAG_sim_b}
\begin{tikzpicture}
\node[draw,text centered] (l) {$L_1$};
\node[right = 1.5 of l, text centered] (a) {$A$};
\node[right=1.5 of a, text centered] (y) {$Y$};
\node[draw,right=1.5 of y,text centered] (l2) {$L_2$};
\draw [->, line width= 1] (a) -- node[above,font=\footnotesize]{}  (y);
\draw [->, line width= 1] (l) -- node[above,font=\footnotesize]{}  (a);
\draw [->, line width= 1] (y) -- node[above,font=\footnotesize]{}  (l2);
\draw[->, line width=1] (l) to  [out=45,in=135, looseness=0.5] node[below,font=\footnotesize]{} (y);
\draw[->, line width=1] (a) to  [out=-45,in=-135, looseness=0.5] node[below,font=\footnotesize]{} (l2);
\end{tikzpicture}
}\\[0.3cm]
\subfloat[DAG for setups 4b and 5]{\label{figure:DAG_sim_c}
\begin{tikzpicture}
\node[draw,text centered] (l) {$L$};
\node[right = 1.5 of l, text centered] (a) {$A$};
\node[right = 1.5 of a, text centered] (y) {$Y$};
\draw [->, line width= 1] (a) -- node[above,font=\footnotesize]{}  (y);
\draw[->, line width=1] (l) to  [out=45,in=135, looseness=0.5] node[below,font=\footnotesize]{} (y);
\end{tikzpicture}
}\hspace*{3cm}
\subfloat[DAG for setup 6]{\label{figure:DAG_sim_d}
\begin{tikzpicture}
\node[text centered] (a) {$A$};
\node[draw,right = 1.5 of t, text centered] (m) {$M$};
\node[right=1.5 of m, text centered] (y) {$Y$};
\draw [->, line width= 1] (a) -- node[above,font=\footnotesize]{}  (m);
\draw [->, line width= 1] (m) -- node[above,font=\footnotesize]{}  (y);
\draw[->, line width=1] (a) to  [out=45,in=135, looseness=0.5] node[below,font=\footnotesize]{} (y);
\end{tikzpicture}
}
\caption{Directed acyclic graphs for simulation setups 1-6; setup 7 is based on Figure \ref{figure:mDAG}.} \label{figure:DAG_sim}
\end{center}
\end{figure}

To illustrate some relevant points that follow from the above sections, it makes sense to look at some simple simulation settings. In total, seven simulation setups are considered. The data-generating mechanisms are visualized in the DAGs from Figures \ref{figure:mDAG} and \ref{figure:DAG_sim}. The exact model specifications are given in the code from Appendix \ref{sec:appendix}.

\begin{itemize}
\item \textit{Setup 1, reference setting with confounding:} The first setting serves as a reference and contains a binary intervention $A$, a continuous confounder $L$ and a continuous outcome $Y$. The relationships between these variables are very simple, see Appendix \ref{sec:appendix} for details. In this setup, a linear regression model (containing the covariates $A$ and $L$) is expected to give an unbiased effect estimate of the average treatment effect.
\item \textit{Setup 2, model mis-specification:} The second setting is identical to the first -- except that the $L$ has a non-linear influence on $Y$, but the regression model includes $L$ linearly. Based on the reasoning from Section \ref{sec:assumptions2}, the effect estimate of $A$ on $Y$ should typically be biased if the model specification (for example with respect to $L$) is incorrect.
\item \textit{Setup 3, inclusion of a collider:} The third setup is similar to the first two settings, but contains an additional variable $L_2$ which is a collider. Inclusion of both the collider and the confounder in the regression model is expected to give biased estimates for the effect of $A$ on $Y$.
\item \textit{Setup 4, effect modification:} The fourth setting is identical to the first, except that the effect of $A$ on $Y$ varies with respect to $L$, i.e. there is effect modification. If the average treatment effect is of interest, it is expected that a linear regression model without an interaction term yields biased effect estimates, whereas the inclusion of an interaction term prohibits the estimation of the marginal target quantity. Setup 4b is based on randomized treatment assignment as illustrated in Figure \ref{figure:DAG_sim_c}. In this case, a regression model containing containing both $A$ and $L$, but not the interaction thereof, should lead to approximately unbiased effect estimates.
\item \textit{Setup 5, collapsibility and conditioning on covariates in randomized experiments:} The fifth setting mimics a randomized experiment because treatment assignment is random and does not depend on $L$, see Figure \ref{figure:DAG_sim_c} and Appendix \ref{sec:appendix}. As opposed to the first four settings, a binary outcome is generated. Given the considerations on collapsibility from Section \ref{sec:assumptions9} it follows that a logistic regression model without inclusion of $L$ should give an unbiased estimate of the \textit{marginal} odds ratio, whereas the inclusion of $L$ in the regression model should yield biased estimates.
\item \textit{Setup 6, conditioning on a mediator:} This setting is based on Figure \ref{figure:DAG_sim_d}. The outcome is normally distributed and both $M$ and $A$ are drawn from a Bernoulli distribution (Appendix \ref{sec:appendix}). A linear model excluding the mediator should give an unbiased effect estimate of the ATE, whereas the inclusion of $M$ in the regression model is expected to yield biased effect estimates.
\item \textit{Setup 7, complete case analysis under a missing not at random setup:} The data of this setting are simulated based on the mechanism explained in Figure \ref{figure:mDAG} and Section \ref{sec:assumptions11}. In this scenario, a complete case analysis should lead to unbiased effect estimates even though the data are missing not at random.
\end{itemize}

\begin{figure}[ht!]
\begin{center}
\includegraphics[width=\textwidth]{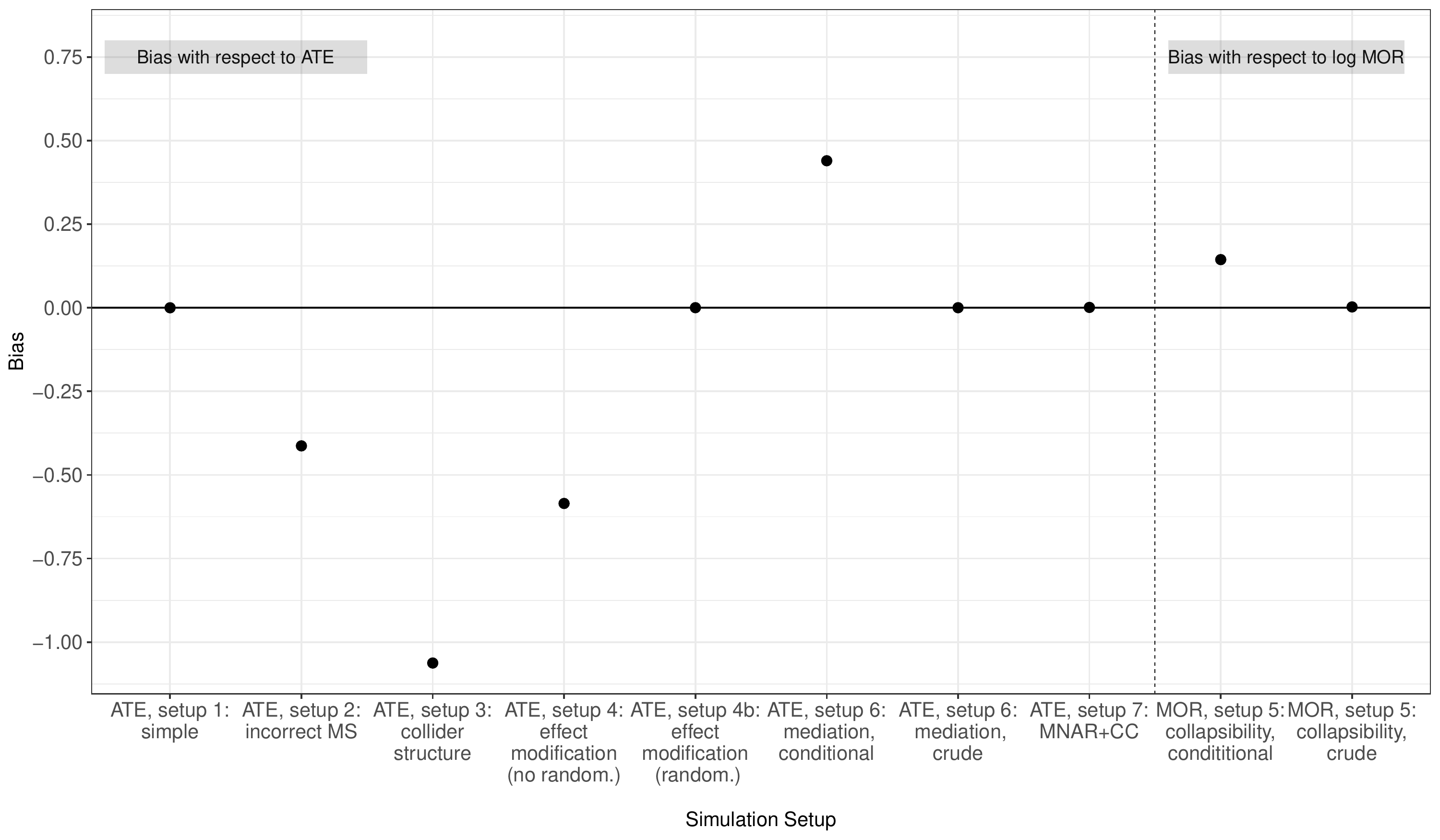}
\caption{Estimated bias in the simulation study for the different scenarios}
\label{figure:simulation}
\end{center}
\end{figure}

The results of the simulation, based on 10.000 simulation runs and a sample size of $n=1000$, are visualized in Figure \ref{figure:simulation}.

As expected, regression can successfully deal with simple confounding (Setup 1). However, if covariates are included but their relationship with the outcome is mis-specified, bias occurs (Setup 2). Inclusion of colliders into a regression model leads to bias too (Setup 3). If effect modification is present, this is irrelevant under randomized treatment assignment (Setup 4b); however, under non-randomized treatment assignment this leads to bias because marginal effect estimation is not possible. When using collapsible effect measures, such as the odds ratio, conditioning on covariates is harmful even in randomized experiments (Setup 5). The results of Setup 6 show that conditioning on mediators can be harmful, whereas Setup 7 illustrates that even under MNAR unbiased effect estimation with regression can be feasible sometimes.

\section{The Implications for Statistical Practice}\label{sec:implications}
Understanding the assumptions which are needed to interpret regression coefficients causally is a prerequisite to understand the differences between different statistical tasks, and how they relate to practical data analyses. A few important points which follow immediately from the above discussions center around the appropriate use and interpretation of regression models, the practice of Monte-Carlo simulations and variable inclusion. These points are discussed in the following sections.

\subsection{Descriptive, Predictive and Explanatory (Causal) Models}\label{sec:implications_trichotomy}
The last decade has seen a number of works which suggest that models, including regression models, should be classified according to the trichotomy ``descriptive -- predictive -- causal'' \cite{Shmueli:2010, Hernan:2019}. The distinction between descriptive and predictive models is sometimes vague, but the difference to models that aim to estimate causal effects is much clearer. Essentially, descriptive and predictive models estimate quantities that can be expressed in terms of summaries of the observed data distribution, whereas causal estimands are defined in terms of the post-intervention distributions, possibly using counterfactual notation. More precisely, regression estimands are often\footnote{this is not always the case because some regression models target conditional densities, such as generalized additive models of location, scale and shape \cite{Stasinopoulos:2017}} conditional expectations, such as $\psi_{F_{\mathcal{O}}} = E(Y|A,\mathbf{L})$, which are a summary of the observed (joint) data distribution $F_{\mathcal{O}}$; however, as explained in Section \ref{sec:framework_target_parameter}, estimands that are of interest from a causal perspective are defined on (joint) post-intervention distributions, such as $E_{F_{\mathcal{P}}}(Y^a)$.

While the estimands for descriptive and predictive questions can be defined based on the observed data distributions, there may still be important differences in terms of estimation: for many descriptive tasks unbiased or consistent estimation may be relevant\footnote{with the exception of data description using multivariate statistics, such as cluster analysis or principal component analysis}; for example, to estimate proportions of unemployment or incidences of a disease, stratified by age and sex. Statistical prediction of new or future observations\footnote{or the estimation of predictive distributions or regions} can be guided by minimizing the prediction error between the estimated predictions and the actual values. Many estimation techniques may reduce particular prediction errors, such as the mean squared prediction error, by introducing bias in exchange for reduced variability: for example, shrinkage estimation \cite{Hastie:2008}, model averaging techniques \cite{Schomaker:2020b, Hansen:2008} or variable selection based on (generalized) cross validation\footnote{the process of variable selection itself produces bias, see Section \ref{sec:implications_variable_selection}}.

Regression models can be used for both description and prediction. The regression coefficients, and their related standard errors, can for example be used to calculate conditional means $E(Y|A,\mathbf{L})$, together with 95\% confidence intervals. They may serve as a purely descriptive summary of the data. If regression models are used for prediction, estimated regression coefficients are part of a function that maps input (covariates) to an output (outcome). However, they can not necessarily be interpreted without making additional causal assumptions about the data generating mechanism. This follows immediately from Section \ref{sec:assumptions}: covariates which are included to minimize a prediction error may be colliders or mediators, which makes their respective coefficients not interpretable; consequently, the confidence intervals related to these regression coefficients may have no meaning too. However, if one commits to a structural model and the assumptions from Section \ref{sec:assumptions} are met, then a regression coefficient can have a causal meaning and may thus be of use for explanation of phenomena.

The practical implications of the above considerations are as follows:
\begin{enumerate}[i)]
\item regression coefficients can only be used for explanation under a causal framework, which requires commitment to a structural model, a target quantity, and evaluation of assumptions 1-12 as outlined above.
\item outside a causal framework regression models may be useful for prediction, but there is no guarantee that regression coefficients and their associated confidence intervals have any (explanatory) meaning.
\item for purely descriptive tasks, regression coefficients may be helpful to calculate conditional expectations (or densities), but following the arguments from ii), the regression coefficients can not be used for explanation, unless causal considerations are being taken into account.
\end{enumerate}

\subsection{Randomized Trials}\label{sec:implications_randomized_trials}
The results and illustrations from Sections \ref{sec:assumptions} and \ref{sec:illustration} have made it clear that non-collapsible effect measures can cause problems even in randomized studies. Even in an idealized experiment, with full compliance, no drop-out and no other problems, the inclusion of additional covariates in the regression models can cause bias with respect to the marginal target quantity. For example, adjusting for variables other than the intervention variable in logistic regression or survival models (such as the Cox proportional hazards model) is incorrect when estimating treatment effects -- unless the inclusion of these variables is meant to evaluate a particular conditional target parameter, like under effect modification hypotheses.

The discussions of this paper have emphasized that the design of randomization alone does not guarantee consistent or unbiased effect estimation under non-compliance, missing data and measurement error.  Non-compliance or non-adherence may be a particular concern in randomized trials: depending on the process that determines non-compliance, using regression can yield biased effect estimates with
respect to the ATE and other common marginal effect measures. As discussed above, re-defining the estimand may be an option in some cases; though accounting for non-adherence by measuring appropriate data and using methods other than regression is the most viable option in most cases \cite{Hernan:2017, Murray:2020}.

Missing data, drop-out and loss to follow-up are not only common for observational data analyses, but also for randomized trials. Both classic and modern results in the missing data literature point towards the fact that neglecting missing data by analyzing the complete cases only will often, though not always, lead to biased effect estimates when using regression and the data are not missing completely at random. Using regression in conjunction with multiple imputation can facilitate consistent effect estimation under assumptions, including a missing at random assumption. As pointed out in the literature, it is however often difficult to defend this assumption when multiple variables are missing. To describe the (causal) missingness process using m-DAGs, and establish identification results for regression estimators based on this, may become more popular in future. A relevant consideration for randomized studies is that censoring study participants due to drop-out may often be informative and the use of survival regression models (without additional modifications or corrections) may not always be correct. Understanding the reasons for missing data is thus certainly crucial in trials.

As discussed above, there are many more consequences for using regression in randomized trials: measurement error will often lead to biased regression coefficients, except in either very
simple situations or situations where mis-measured variables act as a confounder that can block a back-door path. Moreover, even collider bias can be an issue, for example when evaluating (secondary) outcomes on a subset of survivors.

\subsection{Monte-Carlo Simulations}\label{sec:implications_simulations}
Despite the fact that Monte-Carlo simulations are nowadays key to many statistical publications, guidelines and recommendations on best practice are rare, but exist \cite{Morris:2019}. A key point that is often overlooked relates to the fact that any data-generating mechanism follows a natural time-ordering: variables are typically not measured at \textit{exactly} the same time point and observed data associations are the results of an underlying causal data-generating mechanism. This is the case even for survey research: even when two questions are asked shortly after each other, the mechanism that gives rise to the data (i.e. the answers given) are based on attitudes and features that correspond to causal mechanisms; for example, a survey respondent may be asked about age, sex and agreement with a statement addressing discrimination. Simulating data for these variables would follow the natural order of sex, age and agreement with the statement: first, because the response about the statement may be partly determined by age and sex; and (observed) age may relate to sex because the sex distribution in a study population may affect the age distribution as women have another life expectancy than men. Such a time-ordering is clearer for other research areas, such as medicine when pre-treatment variables are followed by treatment and the outcome respectively. It follows that simulation of data which does not resemble some sort of realistic data-generating mechanism is meaningless because any real-world data analysis, upon which statistical methods are employed, is based on such a mechanism. Interestingly, statistical simulations are often based on data-generating mechanisms which are inappropriate. To better understand this, it is useful to compare different practices. Suppose, we have a simple case of 3 variables. One can then distinguish at least four different possibilities of simulating the data-generating process (DGP), all of which are commonly employed in the literature to evaluate estimation strategies in regression models:

\begin{enumerate}
\item Drawing from 3 normal distributions, where $A$ and $L$ are drawn independently of each other:
\begin{lstlisting}
# Simulated data assuming i) independence of variables and
#                        ii) no temporal order
# DGP not useful to evaluate causal/explanatory questions
L <- rnorm(1000,0,1)
A <- rnorm(1000,0,1)
Y <- rnorm(1000,2 - 3*L + 1*A,1)
\end{lstlisting}
\vspace*{-0.7cm}This approach essentially assumes independence between $A$ and $L$ and no temporal ordering\footnote{as written in the code box, one could speculate that $L$ was drawn before $A$, but does not affect $A$ -- and that a time-ordering is present. The first code box means to represent a case where this is \textit{not} assumed however. The fourth box lists the case where an ordering is assumed.}. Because of the latter it is not possible to evaluate post-intervention distributions, and therefore the success of regression in terms of causal effect estimation.
\item Drawing from multivariate distributions:
\begin{lstlisting}
# Simulated data assuming i) dependent variables,
#                        ii) correlation via a multivariate normal distribution
#                       iii) no temporal order
# DGP not useful to evaluate causal/explanatory questions
library(mvtnorm)
sigma <- matrix(c(4,2,2,3), ncol=2)
LA <- rmvnorm(n=1000, mean=c(0,0), sigma=sigma) # matrix of L and A
Y <- rnorm(1000,2 - 3*LA[,1] + 1*LA[,2],1)
\end{lstlisting}
\vspace*{-0.7cm}This approach assumes dependence between $A$ and $L$ and no temporal ordering\footnote{assuming a multivariate normal distribution implies that the respective regression coefficients are determined \cite{Goodnight:1979}; however, this does in general not equate to a particular time-ordering, i.e. the coefficients do not necessarily imply whether A or L is observed first, and thus which variables can affect each other after intervention assignment.}. The dependence structure is defined by, and restricted to, the multivariate normal distribution. As above, it is not possible to evaluate post-intervention distributions, and therefore the success of regression in terms of causal effect estimation.
\item Using copulas\footnote{see Yan (2007) \cite{Yan:2007} for details about the $R$-package \texttt{copula}} to model (possibly non-linear) dependencies between covariates:
\begin{lstlisting}
# Simulated data assuming i) dependent variables,
#                        ii) correlation via a Clayton copula
#                       iii) no temporal order
# DGP not useful to evaluate causal/explanatory questions
library(copula)
mycopula <-  mvdc(claytonCopula(1, dim=2),c("norm","lnorm"),
                  list(list(mean=0.5, sd=1),list(meanlog=0.5, sdlog=0.5)))
LA       <- rMvdc(1000,mycopula)
Y <- rnorm(1000,2 - 3*LA[,1] + 1*LA[,2],1)
\end{lstlisting}
\vspace*{-0.7cm}This approach facilitates possibly complex dependence structures between $A$ and $L$, and assumes no temporal ordering. As above, it is not possible to evaluate post-intervention distributions, and therefore the success of regression in terms of causal effect estimation.
\item Drawing from distributions according to a temporal order:
\begin{lstlisting}
# Simulated data assuming i) dependent variables,
#                        ii) correlation defined by structural relationships
#                       iii) temporal order
# DGP useful to evaluate causal/explanatory questions
L <- rnorm(1000,0,1)
A <- rnorm(1000,2*L,1)
Y <- rnorm(1000,2 + A + 3*L,1)
\end{lstlisting}
\vspace*{-0.7cm}This approach can alternatively also be implemented using the $R$-package \texttt{simcausal} \cite{Sofrygin:2016} which allows technically simple evaluation of causal estimands. Appendix \ref{sec:appendix} gives thorough simulation examples  for the situations considered in Section \ref{sec:illustration}.
\end{enumerate}

The first three approaches explained above have in common that they don't assume any temporal ordering of the variables, or at least the covariates. Since the order is not clear, it is also not clear how intervening on a variable would influence other variables. It is therefore not possible to determine a post-intervention distribution and ultimately causal estimands.

The practical consequences are as follows: if methods of estimation for regression models are evaluated in simulation settings, it is essential to distinguish for what they are being evaluated for: is it a descriptive, predictive or explanatory (causal) question, see also Section \ref{sec:implications_trichotomy}. For descriptive and predictive questions, data-generating processes similar to strategies 1-3 can be appropriate, particularly those that allow for a (potentially complex) dependence structure between variables. In this case, it is particularly meaningful to evaluate measures such as the mean squared predication error, or the bias -- as those are the measures that possibly address predictive and descriptive questions respectively. As argued above, when evaluating estimation techniques for regression coefficients in the context of causal effect estimation, using a time-ordered approach, such as outlined in strategy 4, is inevitable; otherwise, causal concepts such as bias due to conditioning on colliders and mediators, are being neglected implicitly.

\subsection{Variable Selection}\label{sec:implications_variable_selection}
Variable selection in regression models is a highly relevant topic in the contemporary statistical literature. Selecting variables based on hypothesis testing, information criteria (Akaike's information criteria and variations thereof), predictive criteria such as (generalized) cross validation or the area under the curve (AUC), Bayesian criteria, minimum description length and shrinkage estimation have been suggested \cite{Hastie:2008,Rao:2001}. Common arguments to apply model selection relate to a good bias-variance tradeoff, to make a model interpretable by dimension reduction and to find the most important ``predictors'' or ``risk factors''. It is true indeed that model selection often reduces prediction errors: this is essentially because the reduction of included variables reduces variability, in exchange for bias that comes with any kind of model selection\footnote{may it be intended, as for shrinkage estimators such as the LASSO or model averaging techniques\cite{Schomaker:2020b}, or as a byproduct of model selection in general \cite{Hjort:2003}}. Similar to the arguments of Section \ref{sec:implications_trichotomy}, one can conclude that variable selection in regression modeling may be useful for predictive modeling, but the regression coefficients and their respective confidence intervals may not necessarily have a meaning or interpretation: both because colliders or mediators are potentially being selected by the respective model selection procedure and because regression estimates are conditional on the selected model, which results in biased coefficients and over-optimistic confidence intervals \cite{Hjort:2003, Draper:1995, Schomaker:2014}. Of note, the latter point can't always be fixed with post-model selection inference procedures \cite{Leeb:2005, Leeb:2008, Leeb:2008b}.

Variable selection to improve interpretability or to understand ``risk factors'' can only be utilized under a causal framework: as argued above, regression coefficients are only guaranteed to have an explanatory meaning under a causal framework with a clear commitment to a structural causal model, a target quantity and verification and discussion of assumptions 1-12. In this case, variable selection should be solely guided by Pearl's back-door criterion, i.e. inclusion of variables that block back-door paths from the intervention to the outcome and are non-descendants on the intervention.

The ultimate conclusion of these reflections is that it is typically not possible to have one regression model to answer multiple different questions, i.e. to evaluate different treatment effects, make good predictions and describe populations; ideally there is one model per research question, substantiated by structural knowledge if the aim is causal in nature.

\section{Conclusion}\label{sec:conclusions}

This article has highlighted that causal inference with regression modeling is possible, but requires many assumptions to be fulfilled. Many of these assumptions relate directly to the back-door criterion, which requires the inclusion of those variables which close all back-door paths from the intervention to the outcome and no conditioning on descendants of the intervention variable\footnote{of note, the latter point is \textit{not} equivalent to the rule of not conditioning on any post-treatment variable because post-treatment variables may not necessarily be descendants of the intervention, see particularly Section \ref{sec:assumptions4} and Figure \ref{figure:DAG_med_b}}. Additional assumptions, which are not necessarily needed for causal identification in general, but are relevant when estimating and causally interpreting regression models in particular data analyses are consistency (well-defined interventions, non-interference), positivity and considerations regarding the marginal nature of most causal estimands (non-collapsibility, effect modification). Moreover, assumptions regarding the missing data \textit{mechanism}, measurement error and compliance need to be fulfilled.

It is important to stress that some, though not all of these assumptions can be violated even in randomized studies. Using regression in the context of non-collapsible effect measures is a threat which may be less well-known, but potentially problematic in many settings. Also non-compliance with treatment assignment and missing data due to drop out are important considerations that need attention and discussion as described in Sections \ref{sec:assumptions10} and \ref{sec:assumptions11}.

Reflecting on the implications that causal considerations have on estimating and interpreting regression models are manifold: first, without committing to a structural (causal) model and a target quantity, it is impossible to assign any explanatory meaning to a regression coefficient. It follows that the use of a causal framework is indispensable when using regression, unless the scientific question of interest is purely predictive or descriptive. In this case however, there is no guarantee that regression coefficients and their respective confidence intervals have any explanatory meaning (because inclusion of colliders, mediators, incorrect model specification or collapsibility issues may potentially produce estimates that are different in size or sign compared to the [marginal] target quantity of interest); regression coefficients are then either part of a function that maps an input to an output, or a parameter that defines a conditional expectation. This should be taken into account in statistical practice. Second, and related to this point: data-driven model selection can then only be meaningful for predictive tasks; in which case regression coefficients should not necessarily be interpreted (causally) -- see also Section \ref{sec:implications_variable_selection}. Variable inclusion for explanatory questions should be solely guided by the back-door theorem.

Third, as argued above, mechanisms that produce data are naturally causal in the sense that a natural time-ordering and physical laws are the basis of all measurable data. Monte-Carlo simulations which don't take a natural time-ordering into account are often blind to causal concepts related to colliders or mediators; thus, if estimators for regression models are being evaluated by means of simulations, their intended meaning (causal, prediction, description) should be clarified; and if the potential application is explanatory in nature only appropriate data-generating mechanisms, es explained in Section \ref{sec:implications_simulations}, should be used.

Many of the limitations of regression modeling for causal effect estimation can be overcome with competing methods: for example, any issues of conditional versus marginal estimation can be completely avoided by using g-methods, i.e. inverse probability of treatment weighting, g-computation or g-estimation of structural nested models \cite{Hernan:2020, Daniel:2013}. Those methods are sometimes also a good starting point to address issues around non-compliance, missing data and measurement error. To avoid the threat of model mis-specification, it has been suggested to use data adaptive estimation, such as ensemble learning. Doubly robust estimation techniques like targeted maximum likelihood estimation, which require models for both the outcome and treatment assignment process, allow the incorporation of data-adaptive estimation while retaining valid inference \cite{Luque:2018, vanderLaan:2011}. Moreover, flexible and data-adaptive modeling approaches are vital to address practical positivity violations by extrapolating into sparse data regions \cite{Petersen:2012, Schomaker:2019}.

This manuscript did not embark upon simultaneous interventions \cite{Taubman:2009} and longitudinal causal data analyses \cite{Hernan:2020, Daniel:2013}, including multiple time point interventions. However, most relevant points discussed above would also apply in this setting. Moreover, if there are (time-dependent) confounders which are affected by prior interventions, regression estimates becomes invalid anyway \cite{Hernan:2020}.

In summary, regression modeling can be used for causal effect estimation, but it requires many assumptions to be met and practical challenges to be faced, some of which can be avoided using alternative methodological approaches.

\subsection*{Acknowledgements}
I thank Christiane Didden, Miguel Angel Luque-Fernandez, Christian Heumann and Clemence Leyrat for useful discussions of this manuscript.

\bibliographystyle{hunsrt}
{\footnotesize
\bibliography{literature}
}

\clearpage
\appendix

\section{R-Code of the Simulation studies}\label{sec:appendix}

\begin{lstlisting}
####################
# 0) Load Packages #
####################

library(simcausal)  # causal data simulation
library(doParallel) # for parallelization


#################################
# 1) Data-Generating Processes  #
#################################

# a) simple reference setting
M <- DAG.empty()
M <- M +
  node("L",
       distr = "rnorm",
       mean=1, sd=1) +
  node("A",
       distr = "rbern",
       prob = plogis(-0.5 + 2*L)) +
  node("Y",
       distr = "rnorm",
      mean= 2 + A + 3*L)
Mset <- set.DAG(M)   # setup 1
true1 <- 1 # true effect for A

# b) setup for incorrect model specification
M2 <- DAG.empty()
M2 <- M2 +
  node("L",
       distr = "rnorm",
       mean=1, sd=1) +
  node("A",
       distr = "rbern",
       prob = plogis(-0.5 + 2*L )) +
  node("Y",
       distr = "rnorm",
      mean= 2 + A + 0.5*L^2)
Mset2 <- set.DAG(M2) # setup 2
true2 <- 1 # true effect for A

# c) setup for collider bias
M3 <- DAG.empty()
M3 <- M3 +
  node("L",
       distr = "rnorm",
       mean=1, sd=1) +
  node("A",
       distr = "rbern",
       prob = plogis(-0.5 + 2*L)) +
  node("Y",
       distr = "rnorm",
      mean= 2 + A + 3*L) +
  node("L2",                 # L2 = collider
       distr = "rnorm",
       mean=Y*A, sd=1)
Mset3 <- set.DAG(M3) # setup 3
true3 <- 1 # true effect for A


# d) effect modification
M4 <- DAG.empty()
M4 <- M4 +
  node("L",
       distr = "rnorm",
       mean=1, sd=1) +
  node("A",
       distr = "rbern",
       prob = plogis(-0.5 + 2*L)) +
  node("Y",
       distr = "rnorm",
      mean= 2 + A + 3*L + A*L)
Mset4 <- set.DAG(M4) # setup 4
# true marginal ATE
a4.1 <- node("A", distr = "rbern", prob = 1)  # set A=1 and A=0 (intervene)
a4.0 <- node("A", distr = "rbern", prob = 0)
# post-intervention DAG
Mset4 <- Mset4 + action("a4.0", nodes = a4.0) + action("a4.1", nodes = a4.1)
int.dat4 <- simcausal::sim(DAG = Mset4, actions = c("a4.1", "a4.0"), n = 1000000, rndseed = 345)
Mset4 <- set.targetE(Mset4, outcome = "Y", param = "a4.1-a4.0")
true4 <- eval.target(Mset4, data = int.dat4)$res # true effect (ATE) for A

# as d); but with randomized treatment assignment
M4b <- DAG.empty()
M4b <- M4b +
  node("L",
       distr = "rnorm",
       mean=1, sd=1) +
  node("A",            # random treatment assignment
       distr = "rbern",
       prob = plogis(-0.5)) +
  node("Y",
       distr = "rnorm",
      mean= 2 + A + 3*L + A*L)
Mset4b <- set.DAG(M4b) # setup 4
# true marginal ATE
Mset4b <- Mset4b + action("a4.0", nodes = a4.0) + action("a4.1", nodes = a4.1)
int.dat4b <- simcausal::sim(DAG = Mset4b, actions = c("a4.1", "a4.0"),
                            n = 1000000, rndseed = 345)
Mset4b <- set.targetE(Mset4b, outcome = "Y", param = "a4.1-a4.0")
true4b <- eval.target(Mset4b, data = int.dat4b)$res # true effect for A

# e) collapsibility
M5 <- DAG.empty()
M5 <- M5 +
  node("L",
       distr = "rnorm",
       mean=1, sd=1) +
  node("A",
       distr = "rbern",
       prob =  0.5) +   # randomized experiment
  node("Y",             # binary outcome
       distr = "rbern",
        prob = plogis(A+L))
Mset5 <- set.DAG(M5) # setup 5
# true marginal effects
a5.1 <- node("A", distr = "rbern", prob = 1); a5.0 <- node("A", distr = "rbern", prob = 0) # set A=1 and A=0 (intervene)
Mset5 <- Mset5 + action("a5.0", nodes = a5.0) + action("a5.1", nodes = a5.1)
int.dat5  <- simcausal::sim(DAG = Mset5, actions = c("a5.1", "a5.0"), n = 1000000, rndseed = 345)
Mset5     <- set.targetE(Mset5, outcome = "Y", param = "a5.1"); true5.1   <- eval.target(Mset5, data = int.dat5)$res # P(Y(had A=1)=1)
Mset5     <- set.targetE(Mset5, outcome = "Y", param = "a5.0"); true5.2   <- eval.target(Mset5, data = int.dat5)$res # P(Y(had A=0)=1)
true5_ATE <- true5.1-true5.2                                  # true ATE
true5_OR <- ((true5.1)/(1-true5.1))/((true5.2)/(1-true5.2))   # true MOR
true5_logOR <- log(((true5.1)/(1-true5.1))/((true5.2)/(1-true5.2))) # true log MOR

# f) mediation
M6 <- DAG.empty()
M6 <- M6 +
  node("A",
       distr = "rbern",
       prob = plogis(-0.5)) +
  node("M",                           # M = mediator
       distr = "rbern",
       prob = plogis(0.5 - 2*A)) +
  node("Y",
       distr = "rnorm",
      mean= 2 + M + A)
Mset6 <- set.DAG(M6) # setup 6
# true marginal effect estimates
a6.1 <- node("A", distr = "rbern", prob = 1)
a6.0 <- node("A", distr = "rbern", prob = 0)
Mset6 <- Mset6 + action("a6.0", nodes = a6.0) + action("a6.1", nodes = a6.1)
int.dat6 <- simcausal::sim(DAG = Mset6, actions = c("a6.1", "a6.0"), n = 1000000, rndseed = 345)
Mset6 <- set.targetE(Mset6, outcome = "Y", param = "a6.1-a6.0")
true6 <- eval.target(Mset6, data = int.dat6)$res  # true ATE

# g)
M7 <- DAG.empty()
M7 <- M7 +
  node("L1",
       distr = "rnorm",
       mean=1, sd=1) +
  node("L2",
       distr = "rnorm",
       mean=-1, sd=1) +
  node("A",
       distr = "rbern",
       prob = plogis(-0.5 + 2*L1 + L2)) +
  node("CA",             # Prob. that A is missing
       distr = "rbern",
       prob = plogis(1.5+0.5*L1 + 0.5*L2)) +
  node("CL2",            # Prob. that L2 is missing
       distr = "rbern",
       prob = plogis(1.5-0.75*L1 + 0.75*A)) +
  node("Y",
       distr = "rnorm",
      mean= 2 + A)  +
  node("CY",             # Prob. that Y is missing
       distr = "rbern",
       prob = plogis(1.5+0.25*L1 + 0.25*L2 + 0.5*A))
Mset7 <- set.DAG(M7) # setup 7
true7 <- 1           # true ATE

####################
# 2) simulation    #
####################

runs <- 10000  # number of simulation runs
N    <- 1000   # sample size

# Simulation loop
cl <- makeCluster(max(detectCores()-1,1)) # use all cores of computer, minus 1
registerDoParallel(cl)                    # start parallelization

sim <- foreach(r = 1:runs, .combine=rbind, .packages="simcausal") %dopar% {

# draw data
simdat   <- sim(DAG = Mset,   n = N, verbose=F)
simdat2  <- sim(DAG = Mset2,  n = N, verbose=F)
simdat3  <- sim(DAG = Mset3,  n = N, verbose=F)
simdat4  <- sim(DAG = Mset4,  n = N, verbose=F)
simdat4b <- sim(DAG = Mset4b, n = N, verbose=F)
simdat5  <- sim(DAG = Mset5,  n = N, verbose=F)
simdat6  <- sim(DAG = Mset6,  n = N, verbose=F)
simdat7  <- sim(DAG = Mset7,  n = N, verbose=F)
# models for the different setups
m1    <- lm(Y~A+L,data=simdat)
m2    <- lm(Y~A+L,data=simdat2)
m3    <- lm(Y~A+L+L2,data=simdat3)
m4    <- lm(Y~A+L,data=simdat4)
m4b   <- lm(Y~A+L,data=simdat4b)
m5    <- glm(Y~A+L,data=simdat5,family=binomial)
m5.2  <- glm(Y~A,data=simdat5,family=binomial)
m6    <- lm(Y~A+M,data=simdat6)
m6.2  <- lm(Y~A,data=simdat6)
m7    <- lm(Y~A+L1+L2,
                  data=simdat7[simdat7$CY==1 & simdat7$CA==1 & simdat7$CL2==1,])
# store treatment estimates
results <- c(coef(m1)[2], coef(m2)[2], coef(m3)[2], coef(m4)[2], coef(m4b)[2],
             coef(m5)[2], coef(m5.2)[2], coef(m6)[2], coef(m6.2)[2],coef(m7)[2])
}

stopCluster(cl)   # stop parallelization

#########################
# 3) Evaluating Results #
#########################

# Bias
truth <- c(true1,true2,true3,true4,true4b,true5_logOR,true5_logOR,true6,true6,true7)
BIAS <- apply(sim,2,mean)-truth
names(BIAS) <- c("ATE, setup 1:\n simple \n","ATE, setup 2:\n incorrect MS \n",
                 "ATE, setup 3:\n collider \n structure \n",
                 "ATE, setup 4:\n effect \n modification \n (no random.) \n",
                 "ATE, setup 4b:\n effect \n modification \n (random.)\n",
                 "MOR, setup 5:\n collapsibility, \n condititional \n",
                 "MOR, setup 5:\n collapsibility, \n crude \n",
                 "ATE, setup 6:\n mediation,\n conditional \n",
                 "ATE, setup 6:\n mediation,\n crude \n",
                 "ATE, setup 7:\n MNAR+CC \n \n")
BIAS

# Visualize Bias
library(ggplot2)
Bias_data <- as.data.frame(matrix(NA, nrow=length(BIAS),ncol=2))
Bias_data[,1] <- BIAS
Bias_data[,2] <- names(BIAS)
colnames(Bias_data) <- c("Bias","Setup")

pdf(file=paste(getwd(),"/BIAS.pdf",sep=""), width=12)
ggplot(as.data.frame(Bias_data), aes(x=Setup,y=Bias,size=I(2.5))) + geom_point() +
       theme_bw() +
       scale_x_discrete("Simulation Setup") +
       scale_y_continuous("Bias", breaks=seq(-1,1,0.25))  +
  theme(axis.title.x = element_text(size=12), axis.text.x = element_text(size=12),axis.title.y = element_text(size=12, angle = 90), axis.text.y = element_text(size=12))+
  geom_vline(aes(xintercept = 8.5), size = .25, linetype = "dashed") +
  geom_hline(aes(yintercept = 0)) +
  annotate("text", x = 1.5, y = 0.75, label = "Bias with respect to ATE") +
  annotate("text", x = 9.5, y = 0.75, label = "Bias with respect to log MOR")+
  annotate("rect", xmin = 0.5, xmax = 2.5, ymin = 0.7, ymax = 0.8,  alpha = .2)  +
  annotate("rect", xmin = 8.6, xmax = 10.4, ymin = 0.7, ymax = 0.8,  alpha = .2)
dev.off()
\end{lstlisting}

\end{document}